\documentclass[aps,prd,floatfix,superscriptaddress,preprintnumbers,showpacs,showkeys]{revtex4}
\usepackage{amssymb}
\usepackage{amsmath}
\usepackage{amsfonts}
\usepackage{epsfig}  
\usepackage{graphicx}
\usepackage{tabularx}
\newcommand{\seq}{\begin{subequations}}
\newcommand{\sen}{\end{subequations}}
\newcommand{\eq}{\begin{eqnarray}}
\newcommand{\en}{\end{eqnarray}}

\def\shiftdown#1{#1\llap{\lower.04ex\hbox{#1}}}

\newcommand{\ra}{\rangle}
\newcommand{\la}{\langle}

\begin{document}

\title{Electromagnetic structure of nucleon and Roper in soft-wall AdS/QCD} 

\author{Thomas Gutsche}
\affiliation{Institut f\"ur Theoretische Physik,
Universit\"at T\"ubingen, 
Kepler Center for Astro and Particle Physics,  
Auf der Morgenstelle 14, D-72076 T\"ubingen, Germany}
\author{Valery E. Lyubovitskij} 
\affiliation{Institut f\"ur Theoretische Physik,
Universit\"at T\"ubingen, 
Kepler Center for Astro and Particle Physics,  
Auf der Morgenstelle 14, D-72076 T\"ubingen, Germany}
\affiliation{Departamento de F\'\i sica y Centro Cient\'\i fico
Tecnol\'ogico de Valpara\'\i so-CCTVal, Universidad T\'ecnica
Federico Santa Mar\'\i a, Casilla 110-V, Valpara\'\i so, Chile}
\affiliation{Department of Physics, Tomsk State University,  
634050 Tomsk, Russia} 
\affiliation{Laboratory of Particle Physics, Tomsk Polytechnic University,
634050 Tomsk, Russia} 
\author{Ivan Schmidt}
\affiliation{Departamento de F\'\i sica y Centro Cient\'\i fico
Tecnol\'ogico de Valpara\'\i so-CCTVal, Universidad T\'ecnica
Federico Santa Mar\'\i a, Casilla 110-V, Valpara\'\i so, Chile}

\date{\today}

\begin{abstract}

We present an improved study of the electromagnetic form factors 
of the nucleon and of the Roper-nucleon transition using an extended 
version of the effective action of soft-wall AdS/QCD. 
We include novel contribution from additional non-minimal terms, 
which do not renormalize the charge and do not change 
the normalization of the corresponding form factors, but the inclusion
of these terms results in an important contribution to the momentum 
dependence of the form factors and helicity amplitudes. 

\end{abstract}



\maketitle

\section{Introduction}

Originally the soft-wall AdS/QCD action for the nucleon was proposed 
in Ref.~\cite{Abidin:2009hr}. It included a term describing the nucleon
confining dynamics and the electromagnetic field, 
and their minimal and non-minimal couplings 
$Q_N = {\rm diag}(1,0)$ (nucleon charge matrix) 
and $\eta_N= {\rm diag}(\eta_p,\eta_n)$ (nucleon matrix 
of anomalous magnetic moments), respectively. The use 
of the non-minimal couplings is essential to generate the Pauli 
spin-flip form factors. Later, in Ref.~\cite{Vega:2010ns},   
this action was used for the calculation of 
generalized parton distributions of the nucleon. 
In Ref.~\cite{Gutsche:2012bp} it was extended to take into 
account higher Fock states in the nucleon and additional couplings 
with the electromagnetic field in consistency with QCD constituent 
counting rules~\cite{Brodsky:1973kr} for the power scaling of 
hadronic form factors at large values of the momentum transfer 
squared in the Euclidean region. 
In Ref.~\cite{Gutsche:2011vb} soft-wall AdS/QCD was developed 
for the description of baryons with adjustable quantum numbers 
$n$, $J$, $L$, and $S$. In another development, 
in Refs.~\cite{Brodsky:2014yha}-\cite{Chakrabarti:2013dda}, 
the nucleon properties have been analyzed using a Hamiltonian 
formalism. However, their calculation of the nucleon electromagnetic 
properties ignored the contribution of the non-minimal coupling 
to the Dirac form factors, and therefore, the analysis done in 
Refs.~\cite{Brodsky:2014yha}-\cite{Chakrabarti:2013dda},  
is in our opinion not fully consistent. In Ref.\cite{Sufian:2016hwn} 
the ideas of Ref.~\cite{Brodsky:2014yha} has been extended by 
the inclusion of higher Fock states in the nucleon, in order to calculate 
nucleon electromagnetic form factors in light-front holographic QCD. 
In this paper the Pauli form factor is again introduced 
by hand, using the overlap of the $L=0$ and $L=1$ nucleon wave function.  
Additionally, the expression for the neutron Dirac form factor 
has been multiplied by hand by a free parameter $r$. 

In a series of papers~\cite{Gutsche:2013zia}-\cite{Gutsche:2016gcd}
we have developed a light-front quark-diquark 
approach for the nucleon structure, describing nucleon parton distributions 
and form factors from a unified point of view. In particular, 
in a recent paper~\cite{Gutsche:2016gcd} we derived nucleon light-front 
wave functions, analytically matching the results of global fits to the 
quark parton distributions in the nucleon at the initial scale
$\mu \sim 1$ GeV. We also showed that the distributions obey the correct
Dokshitzer-Gribov-Lipatov-Altarelli-Parisi evolution~\cite{DGLAP} 
to high scales. 
Using these constrained nucleon wave functions we get a reasonable 
description of data on nucleon electromagnetic form factors.  
We also predict the transverse parton, Wigner and Husimi distributions
from a unified point of view, using our light-front wave functions and
expressing them in terms of the parton distributions $q_v(x)$ and
$\delta q_v(x)$.  In the context of the nucleon form factors it is also 
important to mention a recent paper~\cite{Brodsky:2016uln}, where the 
$\gamma^* \to \rho$ transition form factor has been calculated in 
soft-wall AdS/QCD. Here it was shown that the form factor is  consistent 
with quark counting rules for differential cross sections with single and 
double vector meson production. It scales as $1/\sqrt{Q^2}$ and,  
therefore, it contributes to the electromagnetic form factors of 
the nucleons at subleading order. 

The Roper-nucleon transition form factors and 
helicity amplitudes can be also discussed within this same formalism. 
The Roper resonance was first considered in the 
context of AdS/QCD in Ref.~\cite{deTeramond:2011qp}, 
where the Dirac form factor for the 
electromagnetic nucleon-Roper transition was calculated in
light-front holographic QCD.  Later, in Ref.~\cite{Gutsche:2012wb}, 
the formalism for the study of nucleon resonances in soft-wall 
AdS/QCD has been developed, and the first application for a detailed 
description of Roper-nucleon transition properties 
(form factors, helicity amplitudes and transition charge radii) 
was performed. In Ref.~\cite{Ramalho:2017pyc,Ramalho:2017muv} 
the formalism proposed in~\cite{Gutsche:2012wb} was used, 
with a different set of parameters. An overview of the
application of other theoretical approaches can be found 
in Refs.~\cite{Gutsche:2012wb,Aznauryan:2012ba}. This includes recent novel  
ideas about considering additional degrees of freedom for
this state, such as a molecular nucleon-scalar $\sigma$ meson 
component~\cite{Obukhovsky:2011sc,Aznauryan:2012ec}, for a realistic 
description of current data on Roper electroproduction performed 
by the CLAS Collaboration 
at JLab~\cite{Aznauryan:2009mx}-\cite{Mokeev:2015lda}. 

In the present paper we include additional non-minimal couplings 
of the vector field (dual to the electromagnetic field) with fermions 
(dual to the nucleon and Roper). Such terms do not renormalize the charge, 
but gives an important contribution to the momentum dependence of the nucleon 
and Roper-nucleon transition form factors (helicity amplitudes). 
The inclusion of these terms helps to improve the description of data. 
The paper is organized as follows. 
In Sec.~II we briefly discuss our formalism. 
In Sec.~III we present the analytical calculation and the numerical analysis 
of electromagnetic form factors and helicity amplitudes of the nucleon 
and the Roper. Finally, Sec.~IV contains our summary and conclusions.

\section{Formalism}

In this section we briefly review our approach. 
We start with the underlying action for the study of the nucleon 
$N = (p,n)$ and Roper ${\cal R} = ({\cal R}_p,{\cal R}_n)$  resonance,
extended by the inclusion of photons. It is constructed in terms of the 
5D AdS fields $\psi^N_{\pm,\tau}(x,z)$ and 
$\psi^{\cal R}_{\pm,\tau}(x,z)$, which are duals to the left- and 
right-handed chiral doublets of nucleons (Roper resonances)
${\cal O}^L = (B_1^L, B_2^L)^T$ and ${\cal O}^R = (B_1^R, B_2^R)^T$ 
with $B_1 = p, {\cal R}_p$ and $B_2 = n, {\cal R}_n$. These fields  
are in the fundamental representations of the chiral $SU_L(2)$ 
and $SU_R(2)$ subgroups 
and are 
holographic analogues of the nucleon $N$ and Roper resonance ${\cal R}$, 
respectively. 
The 5D AdS fields $\psi^B_{\pm,\tau}(x,z)$ are products of the left/right 
4D spinor fields 
\eq
\psi^{L/R}_{n=0,1}(x) = \frac{1 \mp \gamma^5}{2} \, \psi_{n=0,1}(x)\,,
\en
with spin $1/2$ and the bulk profiles 
$F^{L/R}_{\tau, n=0,1}(z) = z^2 \, f^{L/R}_{\tau, n=0,1}(z)$  
with twist $\tau$ depending on the holographic (scale) variable $z$: 
\eq
\psi^N_{\pm,\tau}(x,z) &=& \frac{1}{\sqrt{2}} \,
\left[
      \psi^{L}_0(x) \ F^{L/R}_{\tau, 0}(z)
\pm   \psi^{R}_0(x) \ F^{R/L}_{\tau, 0}(z)\right]\,, \nonumber\\
\psi^{\cal R}_{\pm,\tau}(x,z) &=& \frac{1}{\sqrt{2}} \,
\left[
      \psi^{L}_1(x) \ F^{L/R}_{\tau, 1}(z)
\pm   \psi^{R}_1(x) \ F^{R/L}_{\tau, 1}(z)\right]\,,
\en
where
\eq
f^L_{\tau, 0} &=& \sqrt{\frac{2}{\Gamma(\tau)}} \, \kappa^{\tau} \,
z^{\tau - 1/2} \, e^{-\kappa^2 z^2/2}\,, \nonumber\\
f^R_{\tau, 0} &=& \sqrt{\frac{2}{\Gamma(\tau-1)}} \, \kappa^{\tau-1} \,
z^{\tau - 3/2} \, e^{-\kappa^2 z^2/2}\,, \nonumber\\
f^L_{\tau, 1} &=& \sqrt{\frac{2}{\Gamma(\tau+1)}} \, \kappa^{\tau} \,
z^{\tau - 1/2} \, (\tau  - \kappa^2 z^2) \, e^{-\kappa^2 z^2/2}\,, \nonumber\\
f^R_{\tau, 1} &=& \sqrt{\frac{2}{\Gamma(\tau)}} \, \kappa^{\tau-1} \,
z^{\tau - 3/2} \, (\tau - 1 - \kappa^2 z^2) \, e^{-\kappa^2 z^2/2}\,.
\en
Here the nucleon is identified as the ground state with $n=0$
and the Roper resonance as the first radially excited state with $n=1$.
We also include the vector field $V_M(x,z)$, 
dual to the electromagnetic field. We work in the axial gauge
$V_z = 0$ and perform a Fourier transformation of 
the vector field $V_\mu(x,z)$ with respect to the Minkowski coordinate
\eq\label{V_Fourier}
V_\mu(x,z) = \int \frac{d^4q}{(2\pi)^4} e^{-iqx} V_\mu(q) V(q,z)\,. 
\en
We derive an EOM for the vector bulk-to-boundary propagator $V(q,z)$
dual to the $q^2$-dependent electromagnetic current
\eq
\partial_z \biggl( \frac{e^{-\varphi(z)}}{z} \,
\partial_z V(q,z)\biggr) + q^2 \frac{e^{-\varphi(z)}}{z} \, V(q,z) = 0 \,, 
\en 
where $\varphi(z) = \kappa^2 z^2$ is the dilaton field   
with its scale parameter $\kappa$, which is varied from 380 to 500 MeV 
in different fits to hadron data. 

The solution of this equation in terms of the
gamma $\Gamma(n)$ and Tricomi $U(a,b,z)$ functions reads
\eq
\label{VInt_q}
V(q,z) = \Gamma\Big(1 - \frac{q^2}{4\kappa^2}\Big)
\, U\Big(-\frac{q^2}{4\kappa^2},0,\kappa^2 z^2\Big) \,.
\en
In the Euclidean region ($Q^2 = - q^2 > 0$)
it is convenient to use the integral
representation for $V(Q,z)$~\cite{Grigoryan:2007my}
\eq
\label{VInt}
V(Q,z) = \kappa^2 z^2 \int_0^1 \frac{dx}{(1-x)^2}
\, x^a \,
e^{- \kappa^2 z^2 \frac{x}{1-x} }\,,
\en
where $x$ is the light-cone momentum fraction and
$a = Q^2/(4 \kappa^2)$.

The action contains a free part $S_0$, describing the 
confined dynamics of nucleon, Roper and the electromagnetic field 
in AdS space, and an electromagnetic interaction part $S_{\rm int}$ 
with 
\eq
S   &=& S_0 + S_{\rm int}\,, \nonumber\\
S_0 &=& \int d^4x dz \, \sqrt{g} \, e^{-\varphi(z)} \,
\biggl\{ {\cal L}_N(x,z) + {\cal L}_{\cal R}(x,z)
+ {\cal L}_V(x,z)
\biggr\} \,, \nonumber\\
S_{\rm int} &=& \int d^4x dz \, \sqrt{g} \, e^{-\varphi(z)} \,
\biggl\{
{\cal L}_{VNN}(x,z) + {\cal L}_{V{\cal R}{\cal R}}(x,z)
+ {\cal L}_{V{\cal R}N}(x,z)
\biggr\} \,. 
\en
${\cal L}_N$, ${\cal L}_{\cal R}$, ${\cal L}_V(x,z)$ and
${\cal L}_{VNN}(x,z)$, ${\cal L}_{V{\cal RR}}(x,z)$,
${\cal L}_{V{\cal R}N}(x,z)$ are the free and
interaction Lagrangians,
respectively, and  are written as 
\eq
{\cal L}_B(x,z) &=&  \sum\limits_{i=+,-; \,\tau} \, c_\tau^B \,
\bar\psi^B_{i,\tau}(x,z) \, \hat{\cal D}_i(z) \, \psi^B_{i,\tau}(x,z)
\,, \nonumber\\
{\cal L}_V(x,z) &=& - \frac{1}{4} V_{MN}(x,z)V^{MN}(x,z)\,, \nonumber\\
{\cal L}_{VBB}(x,z) &=& \sum\limits_{i=+,-; \tau} \, c_\tau^B \,
\bar\psi^B_{i, \tau}(x,z) \, \hat{\cal V}^B_i(x,z) \,
\psi^B_{i, \tau}(x,z)\,, \nonumber\\
{\cal L}_{V{\cal R}N}(x,z) &=&
\sum\limits_{i=+,-; \,\tau} \, c_\tau^{{\cal R}N} \,
\bar\psi_{i,\tau}^{\cal R}(x,z) \, \hat{\cal V}^{{\cal R}N}_i(x,z) \,
\psi_{i,\tau}^N(x,z) \, + \, {\rm H.c.}\,,
\en
where $B = N, {\cal R}$ and
\eq
\hat{\cal D}_\pm(z) &=&  \frac{i}{2} \Gamma^M
\! \stackrel{\leftrightarrow}{\partial}_{_M} - \frac{i}{8}
\Gamma^M \omega_M^{ab} [\Gamma_a, \Gamma_b]
\, \mp \,  (\mu + U_F(z))\,, \nonumber\\
\hat{\cal V}^H_\pm(x,z)  &=&  Q \, \Gamma^M  V_M(x,z) \, \pm \,
\frac{i}{4} \, \eta_V^H \,  [\Gamma^M, \Gamma^N] \, V_{MN}(x,z) 
\, \pm \, \frac{i}{4} \, \lambda_V^H \, z^2 \, 
[\Gamma^M, \Gamma^N] \, \partial^K\partial_KV_{MN}(x,z)  
\nonumber\\ 
&\pm& \, g_V^H \, \Gamma^M \, i\Gamma^z \, V_M(x,z)
+     \zeta_V^H \, z \, \Gamma^M  \, \partial^N V_{MN}(x,z)
\, \pm \, \xi_V^H \, z \, \Gamma^M \, i\Gamma^z \, \partial^N V_{MN}(x,z)
\,, \quad H = N, {\cal R}, {\cal R}N\,. 
\en
The set of parameters 
$c_\tau^{N}$, $c_\tau^{{\cal R}}$, and $c_\tau^{{\cal R}N}$ 
induce mixing of the contribution of AdS 
fields with different twist dimension. 
In Refs.~\cite{Gutsche:2012bp,Gutsche:2012wb} we
showed that the parameters $c_\tau^B$  are constrained 
by the condition $\sum_\tau \, c_\tau^B = 1$ in order to get
the correct normalization of the kinetic term
$\bar\psi_n(x)i\!\!\not\!\!\partial\psi_n(x)$
of the four-dimensional spinor field. This condition is also
consistent with electromagnetic gauge invariance. 
The couplings 
$\eta_V^H = {\rm diag}(\eta_V^{H_1},\eta_V^{H_2})$, 
$\lambda_V^H = {\rm diag}(\lambda_V^{H_1},\lambda_V^{H_2})$, 
$g_V^H = {\rm diag}(g_V^{H_1},g_V^{H_2})$,
$\zeta_V^H = {\rm diag}(\zeta_V^{H_1},\zeta_V^{H_2})$,
and
$\xi_V^H = {\rm diag}(\xi_V^{H_1},\xi_V^{H_2})$,
where $H_1 = p, {\cal R}_p,               
{\cal R}_pp$ and $H_2 = n, {\cal R}_n, {\cal R}_nn$ are fixed from
the magnetic moments, slopes, and form factors
of both the nucleon and Roper,
while the couplings $c_\tau^{{\cal R}N}$ are fixed from
the normalization of the Roper-nucleon helicity amplitudes.
The terms proportional to the couplings $\lambda_V^H$, $\zeta_V^H$, 
and $\xi_V^H$ express novel nonminimal couplings of the fermions 
with the vector field. 
It does not renormalize the charge and does not change 
the corresponding form factor normalizations, but gives 
an important contribution to the momentum dependence of 
the form factors and helicity amplitudes. 

We use the conformal metric 
$g_{MN} x^M x^N = \epsilon^a_M \epsilon^b_N 
\eta_{ab} \, x^M x^N = (dx_\mu dx^\mu - dz^2)/z^2$;  
$\epsilon^a_M = \delta^a_M/z$ is the              
vielbein; $\sqrt{g} = 1/z^5$. Here $\mu$ is the                              
five-dimensional mass of the spin-$\frac{1}{2}$ AdS                           
fermion $\mu = 3/2 + L$, with $L$ being the orbital angular momentum;      
$U_F(z) = \varphi(z)$ is the dilaton potential; 
$Q = {\rm diag}(1,0)$ is the nucleon (Roper) charge matrix;  
$V_{MN} = \partial_M V_N - \partial_N V_M$ is                            
the stress tensor for the vector field; 
$\omega_M^{ab} = (\delta^a_M \delta^b_z - \delta^b_M \delta^a_z)/z$ is 
the spin connection term; $\sigma^{MN} = [\Gamma^M, \Gamma^N]$ 
is the commutator of the Dirac matrices in AdS space, which are defined as  
$\Gamma^M = \epsilon^M_a \Gamma^a$ and 
$\Gamma^a = (\gamma^\mu, -i \gamma^5)$.   

The nucleon and Roper masses are identified with the
expressions~\cite{Gutsche:2012bp,Gutsche:2012wb}
\eq\label{Matching1}
M_N = 2 \kappa \sum\limits_\tau\, c_\tau^N\, \sqrt{\tau - 1}
\,, \quad\quad  
M_{\cal R} &=& 2 \kappa \sum\limits_\tau\, c_\tau^{\cal R}\, \sqrt{\tau}\,. 
\en
As we mentioned the set of mixing parameters $c_\tau^{N,{\cal R}}$ 
is constrained by the correct normalization of the kinetic term of the
four-dimensional spinor field and by charge conservation as (see detail in 
Ref.~\cite{Gutsche:2012bp}): 
\eq\label{Matching2}
\sum\limits_\tau \, c_\tau^{N,{\cal R}} = 1\,. 
\en 
Baryon form factors are calculated analytically
using bulk profiles of fermion fields
and the bulk-to-boundary propagator $V(Q,z)$ of 
the vector field (see exact expressions in the next section). 
Calculation technique is discussed in detail
in Refs.~\cite{Gutsche:2012bp,Gutsche:2012wb}.

\section{Electromagnetic form factors of nucleon, 
Roper and Roper-nucleon transitions}

The electromagnetic form factors of the nucleon, Roper and 
Roper-nucleon transitions are defined by the following matrix elements, 
due to Lorentz and gauge invariance,
\eq
N \to N: \hspace*{.5cm}
M^\mu(p_1,\lambda_1;p_2,\lambda_2) &=& \bar u_N(p_2,\lambda_2)
\left[ \gamma^\mu \, F_1^N(q^2) \, - \,
i \sigma^{\mu\nu} \frac{q_\nu}{2M_N} \, F_2^N(q^2) \, \right]
u_N(p_1,\lambda_1)\,,\nonumber\\
{\cal R} \to {\cal R}: \hspace*{.5cm}
M^\mu(p_1,\lambda_1;p_2,\lambda_2) &=& \bar u_{\cal R}(p_2,\lambda_2)
\left[ \gamma^\mu \, F_1^{\cal R}(q^2) \, - \,
i \sigma^{\mu\nu} \frac{q_\nu}{2M_{\cal R}} \, F_2^{\cal R}(q^2) \, \right]
u_{\cal R}(p_1,\lambda_1)\,,\\
{\cal R} \to N: \hspace*{.5cm}
M^\mu(p_1,\lambda_1;p_2,\lambda_2) &=& \bar u_N(p_2,\lambda_2)
\left[ \gamma^\mu_\perp \, F_1^{{\cal R}N}(q^2) \, - \,
i \sigma^{\mu\nu} \frac{q_\nu}{M_{\cal R}+M_N} \, F_2^{{\cal R}N}(q^2) 
\, \right] u_{\cal R}(p_1,\lambda_1)\,,\nonumber
\en
where $\quad \gamma^\mu_\perp = \gamma^\mu - q^\mu \not\! q/q^2$\,, 
$q = p_1 - p_2$, and $\lambda_1$, $\lambda_2$, and $\lambda$ are 
the helicities of the initial, final baryon and photon, obeying the relation 
$\lambda_1 = \lambda_2 - \lambda$. 
 
We recall the definitions of the nucleon Sachs form factors
$G_{E/M}^N(Q^2)$ and the electromagnetic radii $\la r^2_{E/M} \ra^N$
in terms of the Dirac $F_1^N(Q^2)$ and Pauli $F_2^N(Q^2)$ form factors
\eq
G_E^N(Q^2) &=& F_1^N(Q^2) - \frac{Q^2}{4M_N^2} F_2^N(Q^2)\,,
\nonumber\\[2mm] 
G_M^N(Q^2) &=& F_1^N(Q^2) + F_2^N(Q^2)\,, \nonumber\\[2mm]
\la r^2_E \ra^N &=& - 6 \, \frac{dG_E^N(Q^2)}{dQ^2}\bigg|_{Q^2 = 0} \,,
\nonumber\\[2mm]
\la r^2_M \ra^N &=&  - \frac{6}{G_M^N(0)} \,
\frac{dG_M^N(Q^2)}{dQ^2}\bigg|_{Q^2 = 0}  \,,
\en
where $G_M^N(0) \equiv \mu_N$ is the nucleon magnetic moment.

Now we introduce the helicity amplitudes
$H_{\lambda_2\lambda}$, which in turn can be
related to the invariant form factors $F_i^{{\cal R}N}$ (see details in
Refs.~\cite{Kadeer:2005aq,Faessler:2009xn,Branz:2010pq,Gutsche:2017wag}.
The pertinent relation is 
\eq
H_{\lambda_2\lambda} = M_\mu(p_1,\lambda_1;p_2,\lambda_2)
\, \epsilon^{\ast \, \mu}(q,\lambda) \,,
\en
where $\epsilon^{\ast \,\mu}(q, \lambda)$ 
is the polarization vector of the outgoing photon. 
A straightforward calculation 
gives~\cite{Kadeer:2005aq,Faessler:2009xn,Branz:2010pq,Gutsche:2017wag} 
\eq
H_{\pm\frac{1}{2}0} = \sqrt{\frac{Q_-}{Q^2}} \,
\left(
F_1^{{\cal R}N} M_+  - F_2^{{\cal R}N} \frac{Q^2}{M_1} \right) \,, 
\quad \
H_{\pm\frac{1}{2}\pm 1} = - \sqrt{2 Q_-} \,
\left( F_1^{{\cal R}N} + F_2^{{\cal R}N} \frac{M_+}{M_1} \right) \,, 
\en
where $M_\pm = M_1 \pm M_2$, $Q_\pm = M_\pm^2 + Q^2$. 

In the case of the Roper-nucleon transition there exists 
the alternative set of helicity amplitudes $(A_{1/2}, S_{1/2})$
related to the set $(H_{\frac{1}{2}0},H_{\frac{1}{2}1})$
by~\cite{Weber:1989fv,Capstick:1994ne,Copley:1972tu,%
Aznauryan:2007ja,Tiator:2008kd}
\eq
A_{1/2} = - b \, H_{\frac{1}{2}1}\,, \quad
S_{1/2} = b \, \frac{|{\bf p}|}{\sqrt{Q^2}} \, H_{\frac{1}{2}0},
\en
where
\eq
|{\bf p}| = \frac{\sqrt{Q_+ Q_-}}{2M_{\cal R}}\,, \quad  
b = \sqrt{\frac{\pi\alpha}{M_+ M_- M_N }}
\en
and $\alpha = 1/137.036$ is the fine-structure constant. 

Expressions for the electromagnetic form factors of the nucleons, 
Roper, and Roper-nucleon transitions are given as follows: 

nucleon-nucleon transition, 
\eq
F_1^p(Q^2) &=& C_1(Q^2) + g_V^p C_2(Q^2) 
+ \eta_V^p C_3(Q^2) + \lambda_V^p C_4(Q^2) 
+ \zeta_V^p C_5(Q^2) + \xi_V^p C_6(Q^2)
\,,\nonumber\\[2mm]
F_1^n(Q^2) &=& g_V^n  C_2(Q^2) 
+ \eta_V^n C_3(Q^2)  + \lambda_V^n C_4(Q^2) 
+ \zeta_V^n C_5(Q^2) + \xi_V^n C_6(Q^2) 
\,,\nonumber\\[2mm]
F_2^p(Q^2) &=& \eta_V^p C_7(Q^2) + \lambda_V^p C_8(Q^2)\,,
\nonumber\\[2mm]
F_2^n(Q^2) &=& \eta_V^n C_7(Q^2) + \lambda_V^n C_8(Q^2)\,.
\en
Roper-nucleon transition, 
\eq
F_1^{{\cal R}_p p}(Q^2) &=& D_1(Q^2) 
+ g_V^{{\cal R}_pp}         D_2(Q^2)
+ \eta_V^{{\cal R}_pp}      D_3(Q^2)
+ \lambda_V^{{\cal R}_pp}   D_4(Q^2)
+ \zeta_V^{{\cal R}_pp}     D_5(Q^2)
+ \xi_V^{{\cal R}_pp}       D_6(Q^2)
\,,\nonumber\\[2mm]
F_1^{{\cal R}_n n}(Q^2) &=& 
g_V^{{\cal R}_nn}           D_2(Q^2)
+ \eta_V^{{\cal R}_nn}      D_3(Q^2)
+ \lambda_V^{{\cal R}_nn}   D_4(Q^2)
+ \zeta_V^{{\cal R}_nn}     D_5(Q^2)
+ \xi_V^{{\cal R}_nn}       D_6(Q^2)
\,,\nonumber\\[2mm]
F_2^{{\cal R}_p p}(Q^2) &=& \eta_V^{{\cal R}_pp}    D_7(Q^2)
                         +  \lambda_V^{{\cal R}_pp} D_8(Q^2)
\,,\nonumber\\[2mm]
F_2^{{\cal R}_n n}(Q^2) &=& \eta_V^{{\cal R}_nn}    D_7(Q^2)
                         +  \lambda_V^{{\cal R}_nn} D_8(Q^2)
\,.
\en
Roper-Roper transition, 
\eq
F_1^{{\cal R}_p}(Q^2) &=& E_1(Q^2) 
+ g_V^{{\cal R}_p}        E_2(Q^2)
+ \eta_V^{{\cal R}_p}     E_3(Q^2)
+ \lambda_V^{{\cal R}_p}  E_4(Q^2)
+ \zeta_V^{{\cal R}_p}    E_5(Q^2)
+ \xi_V^{{\cal R}_p}      E_6(Q^2)
\,,\nonumber\\[2mm]
F_1^{{\cal R}_n}(Q^2) &=&
    g_V^{{\cal R}_n}      E_2(Q^2) 
+ \eta_V^{{\cal R}_n}     E_3(Q^2)
+ \lambda_V^{{\cal R}_n}  E_4(Q^2)
+ \zeta_V^{{\cal R}_n}    E_5(Q^2)
+ \xi_V^{{\cal R}_n}      E_6(Q^2)
\,,\nonumber\\[2mm]
F_2^{{\cal R}_p}(Q^2) &=& \eta_V^{{\cal R}_p}    E_7(Q^2)
                       +  \lambda_V^{{\cal R}_p} E_8(Q^2)
\,,\nonumber\\[2mm]
F_2^{{\cal R}_n}(Q^2) &=& \eta_V^{{\cal R}_n}    E_7(Q^2)
                       +  \lambda_V^{{\cal R}_n} E_8(Q^2)
\,. 
\en
The structure integrals $C_i(Q^2)$, $D_i(Q^2)$, and $E_i(Q^2)$ 
are given by the analytical expressions (see in Appendix). 
All calculated form factors are consistent with QCD constituent 
counting rules~\cite{Brodsky:1973kr} for the power scaling of 
hadronic form factors at large values of the momentum transfer 
squared in the Euclidean region. 

The parameters, which will be used in the numerical evaluations,
are fixed as follows: we use the universal dilaton parameter of  
$\kappa = 383$ MeV, the sets of twist mixing parameters are fixed 
from data on masses of nucleon 
($c_3^N = 1.800$\,, $c_4^N = -1.042$\,, $c_5^N = 0.242$) 
and Roper 
($c_3^{{\cal R}} = 0.820$\,, $c_4^{{\cal R}} = -0.242$\,, 
 $c_5^{{\cal R}} = 0.422$). 
At fixed $\kappa = 383$ MeV and baryon masses $M_N = 938.27$ MeV 
and $M_{{\cal R}} = 1440$ MeV only two parameters from the set of six twist 
mixing parameters are free. E.g., parameters $c_4^{N,{\cal R}}$ and 
$c_5^{N,{\cal R}}$ can be fixed through the parameters $c_3^{N,{\cal R}}$ 
and ratios $M_{N,{\cal R}}$ using the matching conditions~(\ref{Matching1}) 
and~(\ref{Matching2}). 
The parameters $\eta_V^p =0.2988$ and $\eta_V^n =-0.3188$ are analytically
fixed from data on nucleon magnetic moments:
\eq
\eta_V^p = \Big(\frac{\kappa}{M_N}\Big)^2 \ (\mu_p - 1)\,,
\quad
\eta_V^n = \Big(\frac{\kappa}{M_N}\Big)^2 \ \mu_n\,, 
\en 
where $\mu_p = 2.793$ n.m. and $\mu_n = -1.913$ n.m.~\cite{PDG:2016}.  

The set on the nucleon parameters
$g_V^p = -2.001$\,,
$g_V^n =  1.731$\,,
$\zeta_V^p =-0.109$\,,
$\zeta_V^n =0.101$\,, 
$\xi_V^p =-0.166$\,, 
$\xi_V^n =0.174$\,, 
$\lambda_V^p = - 0.0005$\,, and   
$\lambda_V^n =   0.0012$ 
is fixed from data on electromagnetic radii
and form factors of nucleons.
The set of Roper-nucleon parameters 
$c_3^{{\cal R}N} =  0.142$\,, 
$c_4^{{\cal R}N} = -3.942$\,, 
$c_5^{{\cal R}N} =  3.449$\,,  
$g_V^{{\cal R}_pp} = -10.095$\,, 
$\eta_V^{{\cal R}_pp} = -0.551$\,, 
$\zeta_V^{{\cal R}_pp} = 0.020$\,, and 
$\xi_V^{{\cal R}_pp} = -0.770$ 
is fixed from data on Roper-nucleon 
transition data. For simplicity we put $\lambda_V^{{\cal R}_pp} = 0$. 
Our results for quark and nucleon electromagnetic form factors 
are shown in Figs.~1-9. We compare our results with 
data~\cite{Cates:2011pz}-\cite{Lachniet:2008qf} 
and the dipole fit $G_D(Q^2) = 1/(1+Q^2/\Lambda^2)^2$.  
As scale parameter $\Lambda$ we use two values 
$\Lambda = \sqrt{0.71}$ GeV and $\Lambda = \sqrt{0.66}$ GeV, 
corresponding to the root-mean-square (rms) radius 
$r_p = 0.81$fm and $r_p = 0.84$~fm, respectively. 
In particular, in Fig.~1 and 2 we present our results for the 
Dirac and Pauli $u$ (left panel) and $d$ (right panel) quark form factors. 
Here data are taken from Refs.~\cite{Cates:2011pz,Diehl:2013xca}.

In Fig.~3 we display the Dirac proton form factor multiplied by $Q^4$ 
(left panel) and the ratio $Q^2 F_2^p(Q^2)/F_1^p(Q^2)$ (right panel). 
Results for the Dirac neutron form factor multiplied by $Q^4$ (left panel) 
and ratio $\mu_p G_E^p(Q^2)/G_M^p(Q^2)$ in comparison with global Fit I 
and Fit II (right panel) are shown in Fig.~4. We take 
the central values of the results for a global fit of the charge and 
magnetic proton form factors from Ref.~\cite{Puckett:2017flj}:  

Fit I: 
\eq
G_E^p(Q^2) &=& \frac{1 + a_1^E \tau}
{1 + b_1^E \tau  + c_1^E \tau^2  + d_1^E \tau^3}\,, \nonumber\\ 
G_M^p(Q^2) &=& \frac{1 + a_1^M \tau}
{1 + b_1^M \tau  + c_1^M \tau^2  + d_1^M \tau^3}\,,
\en 
where 
\eq 
& &a_1^E = -0.21\,, \quad 
   b_1^E = 12.21\,, \quad 
   c_1^E = 12.6\,, \quad 
   d_1^E = 23.0\,, \nonumber\\
& &a_1^M = 0.058\,, \quad 
   b_1^M = 10.85\,, \quad 
   c_1^M = 19.9\,, \quad 
   d_1^M = 4.4\,, 
\en 
Fit II: 
\eq
G_E^p(Q^2) &=& \frac{1 + a_2^E \tau}
{1 + b_2^E \tau  + c_2^E \tau^2  + d_2^E \tau^3}\,, \nonumber\\ 
G_M^p(Q^2) &=& \frac{1 + a_2^M \tau}
{1 + b_2^M \tau  + c_2^M \tau^2  + d_2^M \tau^3}\,.
\en 
where 
\eq 
& &a_2^E = -0.01\,, \quad 
   b_2^E = 12.16\,, \quad 
   c_2^E = 9.7\,, \quad 
   d_2^E = 37.0\,, \nonumber\\
& &a_2^M = 0.093\,, \quad 
   b_2^M = 11.07\,, \quad 
   c_2^M = 19.1\,, \quad 
   d_2^M = 5.6\,. 
\en  
Here $\tau = Q^2/(4 M_N^2)$. 

In Figs.~5 and 6 we present the ratios $G_E^p(Q^2)/G_D(Q^2)$ and 
$G_M^p(Q^2)/(\mu_p G_D(Q^2))$ in comparison with the global Fit I and Fit II 
for the dipole scale parameter $\Lambda^2 = 0.71$ GeV$^2$ 
(left panel) and $\Lambda^2 = 0.66$ GeV$^2$ (right panel). 
A detailed comparison of different ratios of 
the nucleon Sachs form factors is shown in Fig.~7-9. 
Here we use the dipole function $G_D(Q^2)$ with $\Lambda^2 = 0.71$ GeV$^2$. 
The Roper-nucleon transition form factors and helicity amplitudes 
are shown in Figs.~10 and 11. 
Our predictions for the Roper-nucleon 
helicity amplitudes are compared with 
experimental data of the CLAS (JLab)~\cite{Mokeev:2015lda} and 
A1 (MAMI)~\cite{Stajner:2017fmh} Collaborations, and with 
the MAID parametrization~\cite{Drechsel:2007if} 
\eq
A_{1/2}^p(Q^2) &=& -0.0614 \ {\rm GeV}^{-1/2} \
(1 - 1.22 \ {\rm GeV}^{-2} Q^2 - 0.55 \ {\rm GeV}^{-8} Q^8)
\exp[-1.51 \ {\rm GeV}^{-2} Q^2]\,,
\nonumber\\
S_{1/2}^p(Q^2) &=&
0.0042 \ {\rm GeV}^{-1/2} \
(1 + 40 \ {\rm GeV}^{-2} Q^2 + 1.5 \ {\rm GeV}^{-8} Q^8)
\exp[-1.75 \ {\rm GeV}^{-2} Q^2]\,,
\en
and with the parametrization proposed by us. 
We find that the present data on helicity amplitudes can be fitted 
with the use of the formulas  
\eq
A_{1/2}^p(Q^2) &=& A_{1/2}^p(0) \,
\frac{1 + a_1 Q^2}{1 + a_2 Q^2 + a_3 Q^4 + a_4 Q^6} \,,
\nonumber\\
S_{1/2}^p(Q^2) &=& S_{1/2}^p(0) \,
 \frac{1 + s_1 Q^2}{1 + s_2 Q^2 + s_3 Q^4 + s_4 Q^6} \,,
\en
where 
\eq
A_{1/2}^p(0) = - 0.064 \ {\rm GeV}^{-1/2}\,, \quad
S_{1/2}^p(0) =   0.010\ {\rm GeV}^{-1/2}\,, \quad
\en
and
\eq
& &
a_1 = - 2.03556  \ {\rm GeV}^{-2}\,, \quad
a_2 =   1.24891  \ {\rm GeV}^{-2}\,, \quad
a_3 = - 0.90673  \ {\rm GeV}^{-4}\,, \quad
a_4 =   0.41896  \ {\rm GeV}^{-6}\,,\nonumber\\
& &
s_1 = 16.59500   \ {\rm GeV}^{-2}\,, \quad
s_2 =  1.75908   \ {\rm GeV}^{-2}\,, \quad
s_3 =  3.91487   \ {\rm GeV}^{-4}\,, \quad
s_4 =- 0.15289   \ {\rm GeV}^{-6}\,.
\en 
Our results for magnetic moments, slope radii and 
Roper-nucleon transition helicity amplitudes at $q^2 = 0$ 
are summarized in Table I. 

\begin{table}[ht]
\begin{center}
\caption{Electromagnetic properties of nucleons and Roper}

\vspace*{.1cm}

\def\arraystretch{1.25}
    \begin{tabular}{|c|c|c|}
      \hline
Quantity & Our results & Data~\cite{PDG:2016}                  \\
\hline
$\mu_p$ (in n.m.)          &  2.793       &  2.793              \\
\hline
$\mu_n$ (in n.m.)          & -1.913       & -1.913              \\
\hline
$r_E^p$ (fm)     &  0.832 &  0.84087 $\pm$ 0.00039 \\
                 &        &  0.8751  $\pm$ 0.0061  \\
\hline
$\la r^2_E \ra^n$ (fm$^2$) & -0.116 & -0.1161 $\pm$ 0.0022 \\
\hline
$r_M^p$ (fm)     &  0.793 &  0.78  $\pm$ 0.04 \\
\hline
$r_M^n$ (fm)     &  0.813 &  0.864$^{+0.009}_{-0.008}$     \\
\hline
$A_{1/2}^p(0)$ (GeV$^{-1/2}$) & -0.061 & -0.060 $\pm$ 0.004 \\
\hline
$S_{1/2}^p(0)$ (GeV$^{-1/2}$) &  0.008 & --- \\
\hline
\end{tabular}
\end{center}
\end{table}

\section{Summary}

In the present paper we significantly improved the description of both the 
nucleon and the Roper structure using a soft-wall AdS/QCD approach. 
We included novel contributions to the AdS/QCD action  
from additional non-minimal terms, 
which do not renormalize the charge and do not change 
the normalization of the corresponding form factors. They give 
important contributions to the momentum dependence of 
the form factors and helicity amplitudes in reasonable agreement 
with data. In the future we plan to extend our formalism to the study  of
other nucleon resonances. 

\begin{acknowledgments}

This work was supported 
by the German Bundesministerium f\"ur Bildung und Forschung (BMBF)
under Project 05P2015 - ALICE at High Rate (BMBF-FSP 202):
``Jet- and fragmentation processes at ALICE and the parton structure 
of nuclei and structure of heavy hadrons'', 
by CONICYT (Chile) Research Project No. 80140097
and under Grants No. 7912010025, 1140390 and PIA/Basal FB0821, 
by Tomsk State University Competitiveness
Improvement Program and the Russian Federation program ``Nauka''
(Contract No. 0.1764.GZB.2017), and by Tomsk Polytechnic University 
Competitiveness Enhancement Program (Grant No. VIU-FTI-72/2017).  

\end{acknowledgments} 

\appendix
\section{The structure integrals $C_i(Q^2)$, $D_i(Q^2)$, and $E_i(Q^2)$} 

Functions $C_i(Q^2)$, $D_i(Q^2)$, and $E_i(Q^2)$ are given by the analytical expressions 
\eq
C_{i}(Q^2) &=& \sum\limits_\tau \, c_\tau^N \, C_{i}^\tau(Q^2)\,,
\nonumber\\
C_1^\tau(Q^2) &=&
B(a+1,\tau) \, \left(\tau + \frac{a}{2}\right) \,, \nonumber\\
C_2^\tau(Q^2) &=&
\frac{a}{2} \, B(a+1,\tau) \,, \nonumber\\
C_3^\tau(Q^2) &=& a \, B(a+1,\tau+1) \,
\frac{a (\tau - 1) - 1}{\tau}
\,, \nonumber\\
C_4^\tau(Q^2) &=& 2 \, a \, \biggl[ 
(\tau - 1) \, B(a+1,\tau) \,-\, 
2 (2 \tau - 1) \, B(a+1,\tau+1) \,+\, 
3 (\tau+1) \, B(a+1,\tau+2) \nonumber\\
&+& 2 \, (\tau^2 - 1) \, B(a+2,\tau+1) \,-\,
2 (\tau + 1) (\tau + 2) \, B(a+2,\tau+2) \biggr] 
\,, \nonumber\\
C_5^\tau(Q^2)
&=& - a \, \biggl[
(\tau - 1) \, B(a+1,\tau) \,+\,
\tau (2\tau - 1) \, B(a+1,\tau+1) \,+\,
2\tau (\tau + 1) \, B(a+1,\tau+2)\biggr]\,, \nonumber\\
C_6^\tau(Q^2)
&=& - a \, \biggl[
(\tau - 1) \, B(a+1,\tau) \,+\,
\tau (2\tau - 3) \, B(a+1,\tau+1) \,-\,
2\tau (\tau + 1) \, B(a+1,\tau+2)\biggr]\,, \nonumber\\
C_7^\tau(Q^2) &=& \frac{2 M_N}{\kappa} \,  (a+1+\tau) \, \sqrt{\tau - 1} 
\, B(a+1,\tau+1) \,, \nonumber\\
C_8^\tau(Q^2) &=& \frac{4 M_N}{\kappa} \, a \, \tau \, 
\sqrt{\tau - 1} \, \biggl[ 
B(a+1,\tau+1) \,+\, 2 (\tau + 1) B(a+1,\tau+2) \biggr]\,,  
\en 
\eq 
D_{i}(Q^2) &=& \sum\limits_\tau \, c_\tau^{{\cal R}N} \, D_{i}^\tau(Q^2)\,,
\nonumber\\
D_{1}^\tau(Q^2) &=&
\frac{a}{2} \, B(a+1,\tau+1) \,
\biggl[ \sqrt{\tau-1} \biggl(1 + \frac{a+1}{\tau} \biggr) + \sqrt{\tau}
\biggr] \,, \nonumber\\
D_{2}^\tau(Q^2) &=&
\frac{a}{2} \, B(a+1,\tau+1) \,
\biggl( \sqrt{\tau-1} \biggl(1 + \frac{a+1}{\tau} \biggr) - \sqrt{\tau}
\biggr) \,, \nonumber\\
D_{3}^\tau(Q^2) &=&  a 
\biggl[ (\tau - 1)^{3/2} \, B(a+1,\tau) 
- \tau \, (\sqrt{\tau} + \sqrt{\tau-1}) \, B(a+1,\tau+1) 
+ (\tau + 1) \sqrt{\tau} \, B(a+1,\tau+2) \biggr] 
\,, \nonumber\\
D_{4}^\tau(Q^2) &=& 2 a \biggl[ 
\tau (\tau - 1)^{3/2} \, B(a+1,\tau+1) \, + \,
\tau^{3/2} ((\tau - 2) \sqrt{\tau (\tau-1)}  - \tau - 1) \, B(a+1,\tau+2) 
\nonumber\\
&-& (\tau + 1) \sqrt{\tau (\tau-1)} \,\,  
(2 \, (\tau + 1) \sqrt{\tau-1} + (4 \tau - 1) \sqrt{\tau}\,) \, B(a+1,\tau+3) 
\nonumber\\
&+&  
(\tau + 1) (\tau + 2) \sqrt{\tau} \, (3 + 4\tau + 2 \sqrt{\tau (\tau-1)}) \, 
B(a+1,\tau+4)\nonumber\\
&-&  
2 (\tau + 1) (\tau + 2) (\tau + 3) \sqrt{\tau} \,  
B(a+1,\tau+5) \biggr]\,,  
\nonumber\\
D_{5}^\tau(Q^2) &=& - a \,
\biggl[ (\tau - 1)^{3/2} \, B(a+1,\tau) \, + \,
\tau \, (\sqrt{\tau} + \sqrt{\tau-1} (2\tau - 3)) \, B(a+1,\tau+1)
\nonumber\\
&+&
\sqrt{\tau} \, (\sqrt{\tau} - \sqrt{\tau-1})^2 \, (\tau + 1) \, B(a+1,\tau+2)
- 2 \sqrt{\tau} (\tau + 1) (\tau + 2) \, B(a+1,\tau+3) \biggr]
\,, \nonumber\\
D_{6}^\tau(Q^2) &=& - a \,
\biggl[ (\tau - 1)^{3/2} \, B(a+1,\tau) \, - \,
\tau \, (\sqrt{\tau} - \sqrt{\tau-1} (2\tau - 3)) \, B(a+1,\tau+1)
\nonumber\\
&-&
\sqrt{\tau} \, (\sqrt{\tau} + \sqrt{\tau-1})^2 \, (\tau + 1) \, B(a+1,\tau+2)
+ 2 \sqrt{\tau} (\tau + 1) (\tau + 2) \, B(a+1,\tau+3) \biggr]
\,, \nonumber\\
D_{7}^\tau(Q^2) &=& \frac{M_N+M_{\cal R}}{2 \kappa} \, B(a+1,\tau+1) \,
\biggl[
a(\tau - 1) - \tau - 1 + a \sqrt{\tau (\tau - 1)}
\biggr] \,,\nonumber\\
D_{8}^\tau(Q^2) &=& \frac{M_N+M_{\cal R}}{\kappa} \, a \, 
\sqrt{\tau} \, (\sqrt{\tau} + \sqrt{\tau-1}) 
\biggl[ \sqrt{\tau (\tau-1)} \, B(a+1,\tau+1) 
\nonumber\\
&+& 
(\tau+1) \, (2 \sqrt{\tau (\tau-1)} - 1) \, B(a+1,\tau+2) 
\,-\, 2 (\tau+1) (\tau+2) \, B(a+1,\tau+3) \biggr] \,,
\en
and
\eq
E_{i}(Q^2) &=& \sum\limits_\tau \, c_\tau^{\cal R} \,
E_{i}^\tau(Q^2)\,,
\nonumber\\
E_1^\tau(Q^2) &=&  \frac{1}{2}
\biggl[(\tau - 1)^2 \, B(a+1,\tau-1) + \tau (2 - \tau) B(a+1,\tau)
- \tau (\tau + 1) B(a+1,\tau+1)   \nonumber\\
&+& (\tau + 1) (\tau + 2) B(a+1,\tau+2)\biggr]
\,, \nonumber\\
E_2^\tau(Q^2) &=&  \frac{1}{2}
\biggl[(\tau - 1)^2 \, B(a+1,\tau-1) + \tau (2 - 3\tau) B(a+1,\tau)
\nonumber\\
&+& 3\tau (\tau + 1) B(a+1,\tau+1)
- (\tau + 1) (\tau + 2) B(a+1,\tau+2)\biggr]
\,, \nonumber\\
E_3^\tau(Q^2) &=&  a
\biggl[(\tau - 1)^2 \, B(a+1,\tau) + \tau (2 - 3\tau) B(a+1,\tau+1)
\nonumber\\
&+& 3\tau (\tau + 1) B(a+1,\tau+2)
- (\tau + 1) (\tau + 2) B(a+1,\tau+3)\biggr]
\,, \nonumber\\
E_4^\tau(Q^2) &=&  2 a (\tau+1) \biggl[ 
\tau (\tau-1)^2 \, B(a+1,\tau+2) \,+\, 
\tau (2 \tau^3 - 6 \tau^2 - 4 \tau + 5) \, B(a+1,\tau+3)
\nonumber\\
&-& 
\tau (\tau+2) \, (8 \tau^2 - 2 \tau   - 13) \, B(a+1,\tau+4) \,+\, 
(\tau+2) (\tau+3) \,  (12 \tau^2 + 10 \tau  - 4) \, B(a+1,\tau+5) 
\nonumber\\
&-& 
(8 \tau + 7) (\tau+2) (\tau+3) (\tau+4) \, B(a+1,\tau+6) 
\,+\, 2 (\tau+2) (\tau+3) (\tau+4) (\tau+5) \, B(a+1,\tau+7) \biggr]
\,, \nonumber\\
E_5^\tau(Q^2)
&=& - a \, \biggl[
(\tau - 1) \, B(a+1,\tau) \,+\,
2 \tau (\tau - 1) \, B(a+1,\tau+1)
\nonumber\\
&-& (\tau + 1) \, B(a+1,\tau+2) 
\,-\, 2 (\tau+1) (\tau+2) B(a+1,\tau+3)\biggr]
\,, \nonumber\\
E_6^\tau(Q^2)
&=& - a \, \biggl[
(\tau - 1) \, B(a+1,\tau) \,+\,
2 \tau (\tau - 2) \, B(a+1,\tau+1)
\nonumber\\
&-& (\tau + 1) (4 \tau - 1)\, B(a+1,\tau+2) 
\,+\, 2 (\tau+1) (\tau+2) B(a+1,\tau+3)\biggr]
\,, \nonumber\\
E_7^\tau(Q^2) &=& \frac{2 M_{\cal R}}{\kappa}
\, \sqrt{\tau}
\biggl[\tau (\tau - 1) \, B(a+1,\tau) - (\tau + 1) (2 \tau  - 1)
B(a+1,\tau+1)
\nonumber\\
&+& (\tau + 1) (\tau + 2) B(a+1,\tau+2) \biggr]\,, 
\nonumber\\
E_8^\tau(Q^2) &=& \frac{4 M_{\cal R}}{\kappa} 
\, a \, \sqrt{\tau} (\tau+1) 
\biggl[\tau (\tau - 1) \, B(a+1,\tau+2) \,+\, 
(2 \tau^2 - 4 \tau + 1) (\tau + 2) \, B(a+1,\tau+3) 
\nonumber\\
&+& 
(3 - 4 \tau) (\tau + 2) (\tau + 3) B(a+1,\tau+4) \,+\, 
2 \, (\tau + 2) (\tau + 3) (\tau + 4) B(a+1,\tau+5) \biggr]
\,,
\en
where 
\eq
B(m,n) = \frac{\Gamma (m) \Gamma (n)}{\Gamma (m+n)}
\en
is the Beta function.

\begin{figure}
\begin{center}
\epsfig{figure=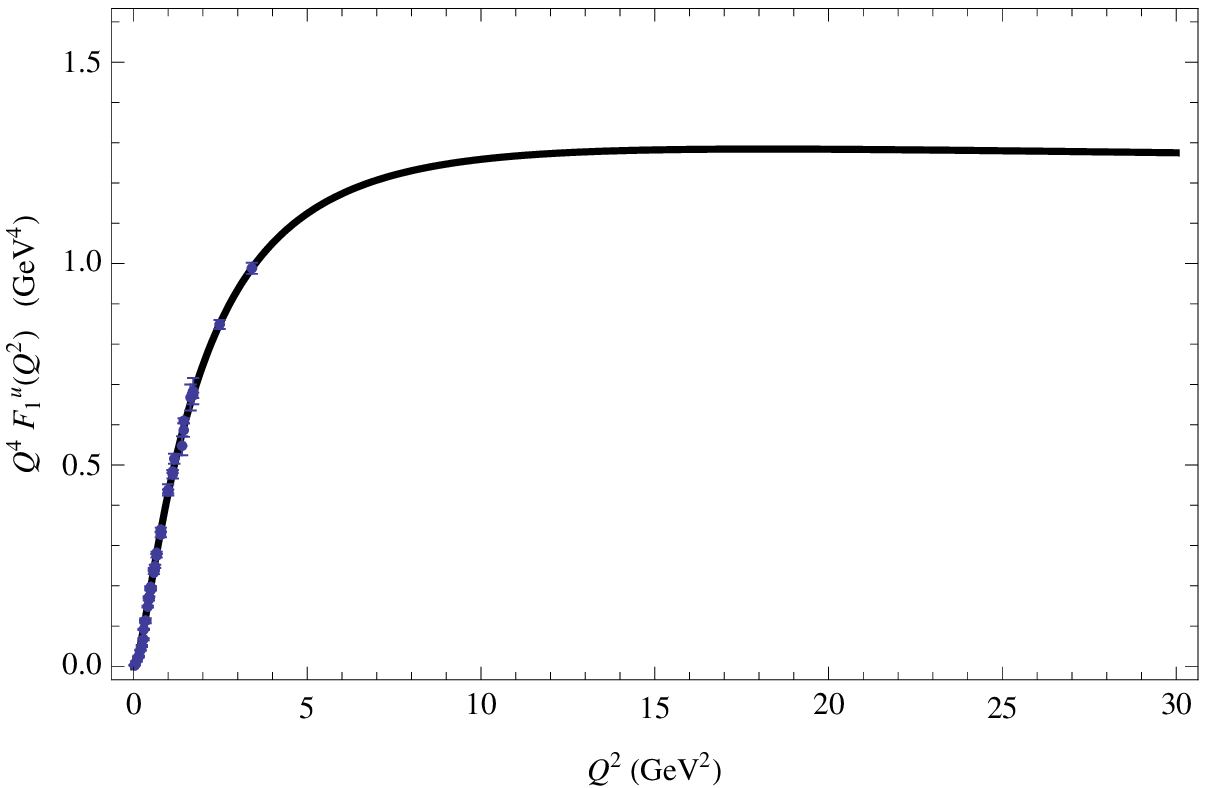,scale=.65}
\epsfig{figure=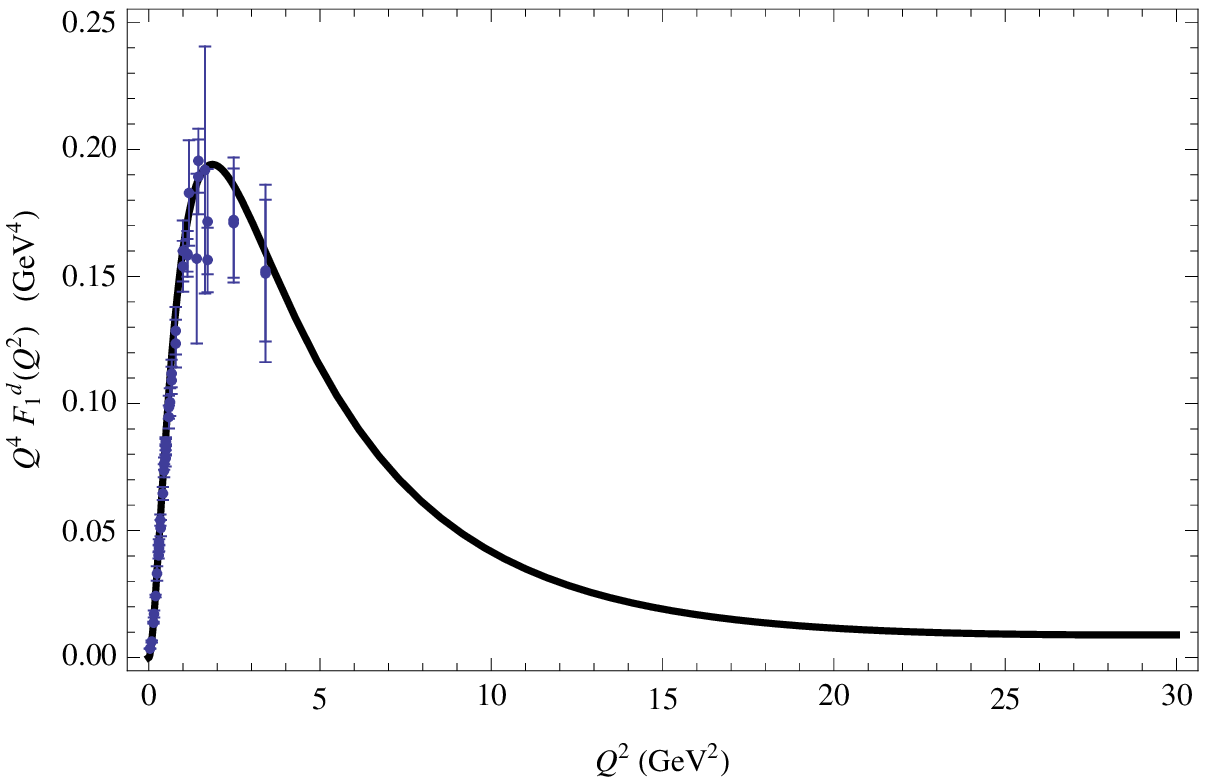,scale=.65}
\end{center}
\noindent
\caption{Dirac $u$ and $d$ quark
form factors multiplied by $Q^4$.
\label{fig1}}

\vspace*{.5cm}
\begin{center}
\epsfig{figure=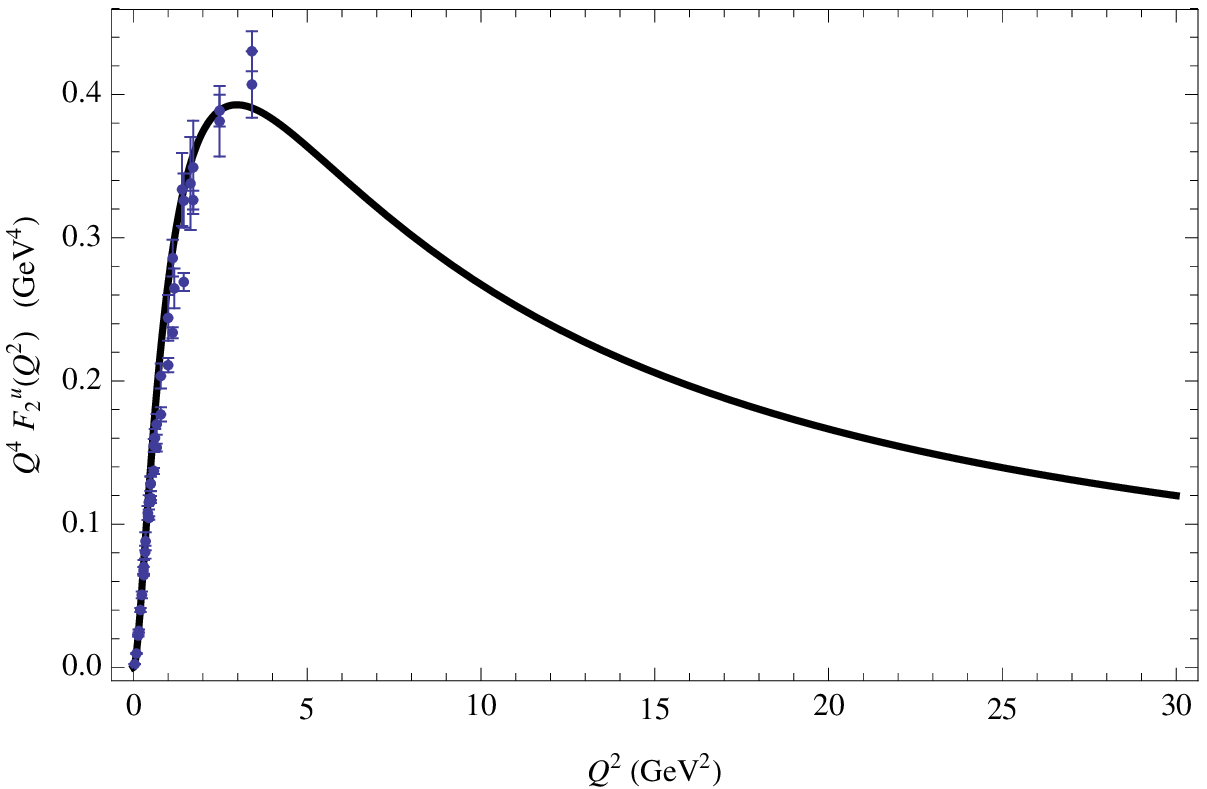,scale=.65}
\epsfig{figure=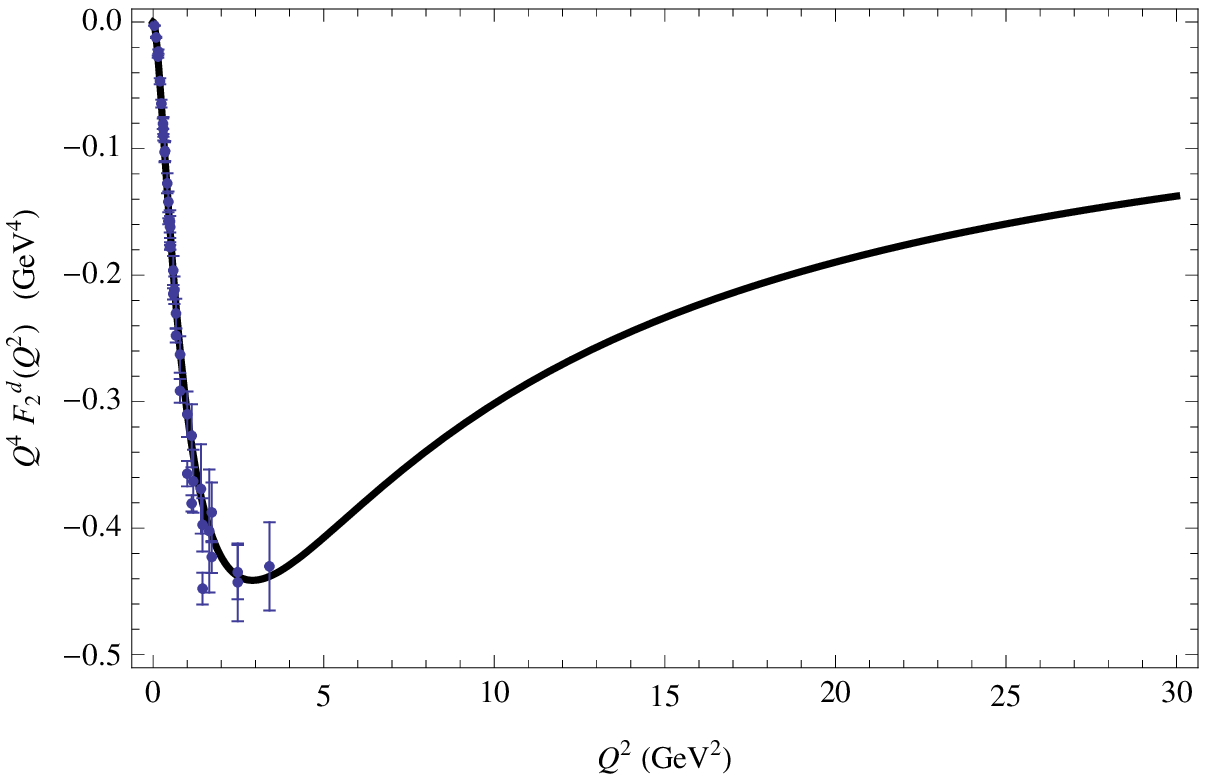,scale=.65}
\end{center}
\noindent
\caption{Pauli $u$ and $d$ quark
form factors multiplied by $Q^4$.
\label{fig2}}

\vspace*{.5cm}
\begin{center}
\epsfig{figure=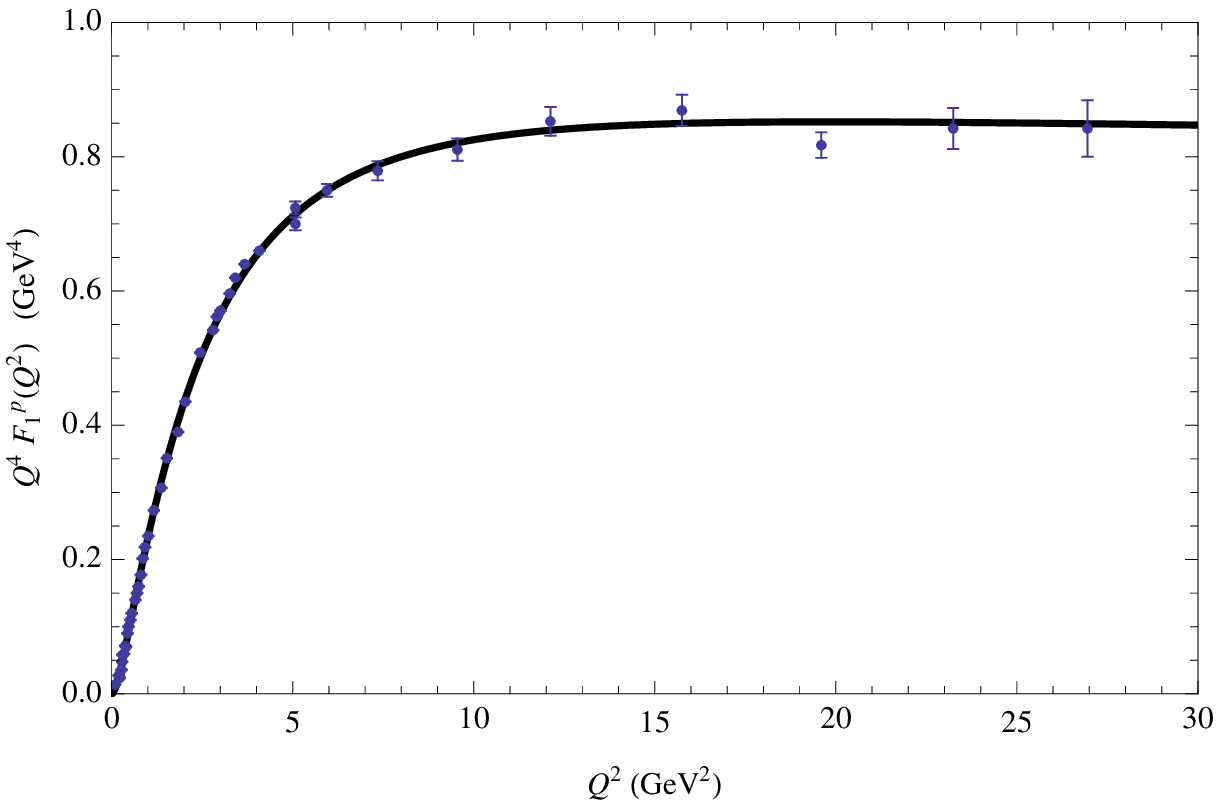,scale=.65}
\epsfig{figure=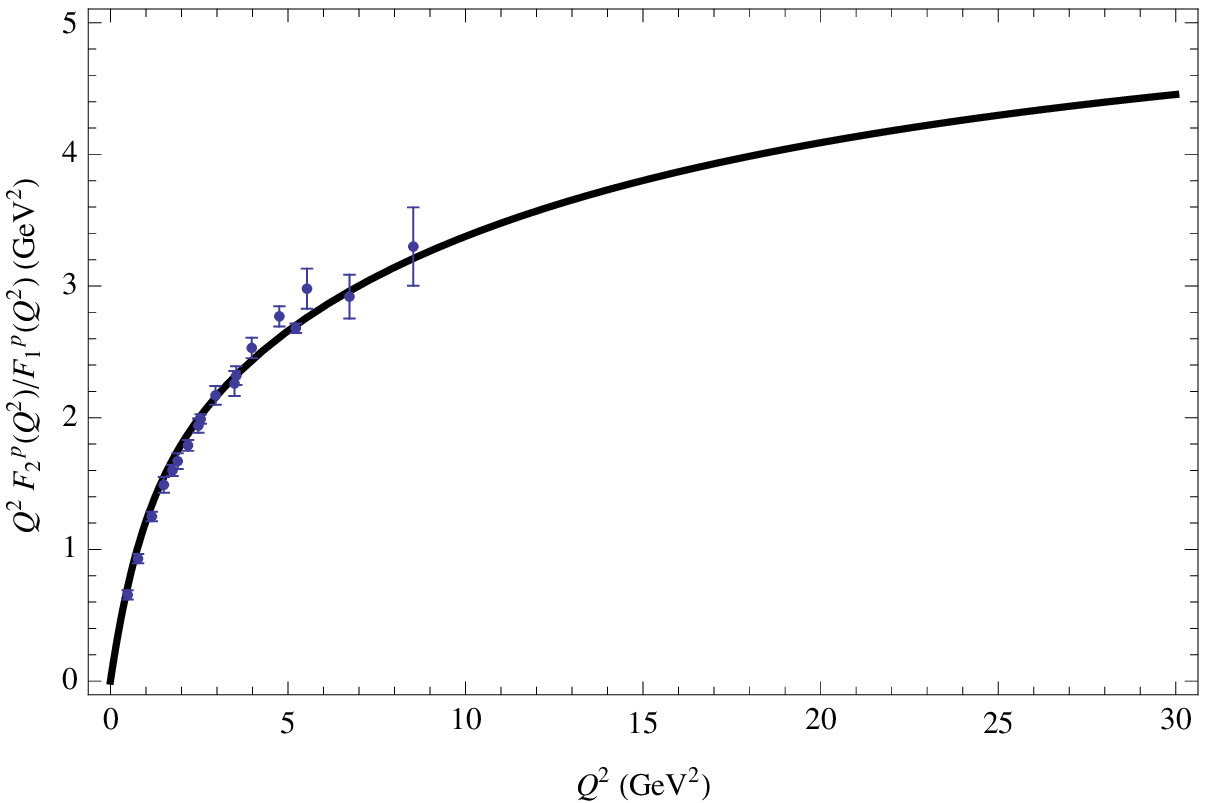,scale=.65}
\end{center}
\noindent
\caption{Dirac proton
form factor multiplied by $Q^4$ 
and ratio $Q^2 F_2^p(Q^2)/F_1^p(Q^2)$.
\label{fig3}}
\end{figure}

\clearpage 

\begin{figure}
\begin{center}
\epsfig{figure=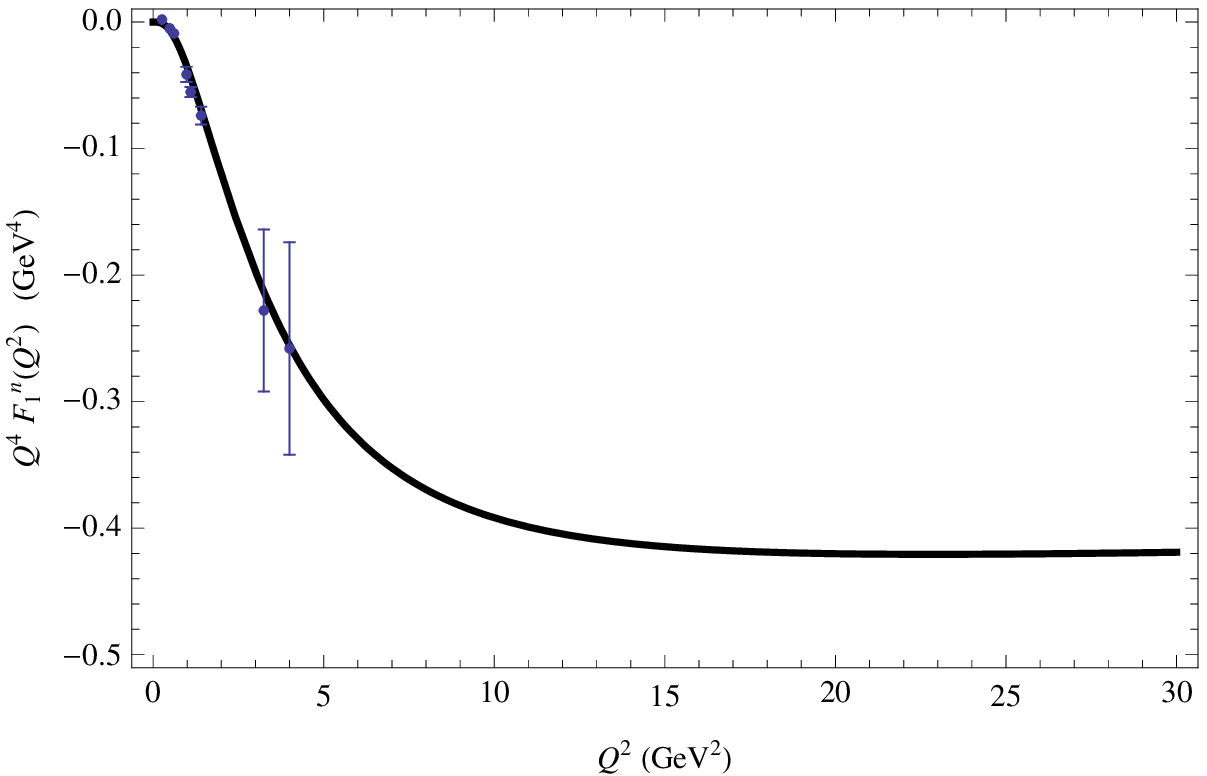,scale=.65}
\epsfig{figure=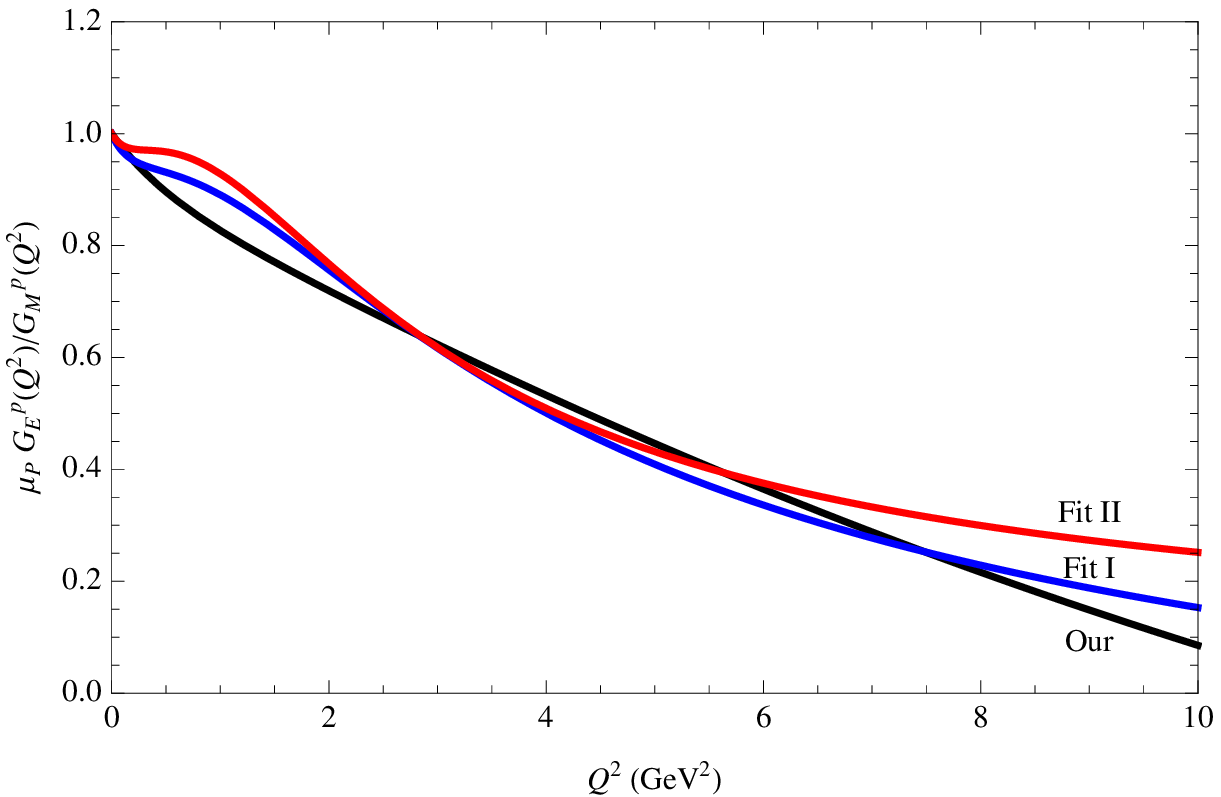,scale=.65}
\end{center}
\noindent
\caption{Dirac neutron 
form factor multiplied by $Q^4$ 
and ratio $\mu_p G_E^p(Q^2)/G_M^p(Q^2)$ 
in comparison with global Fit I and Fit II.    
\label{fig4}}

\vspace*{.5cm}

\begin{center}
\epsfig{figure=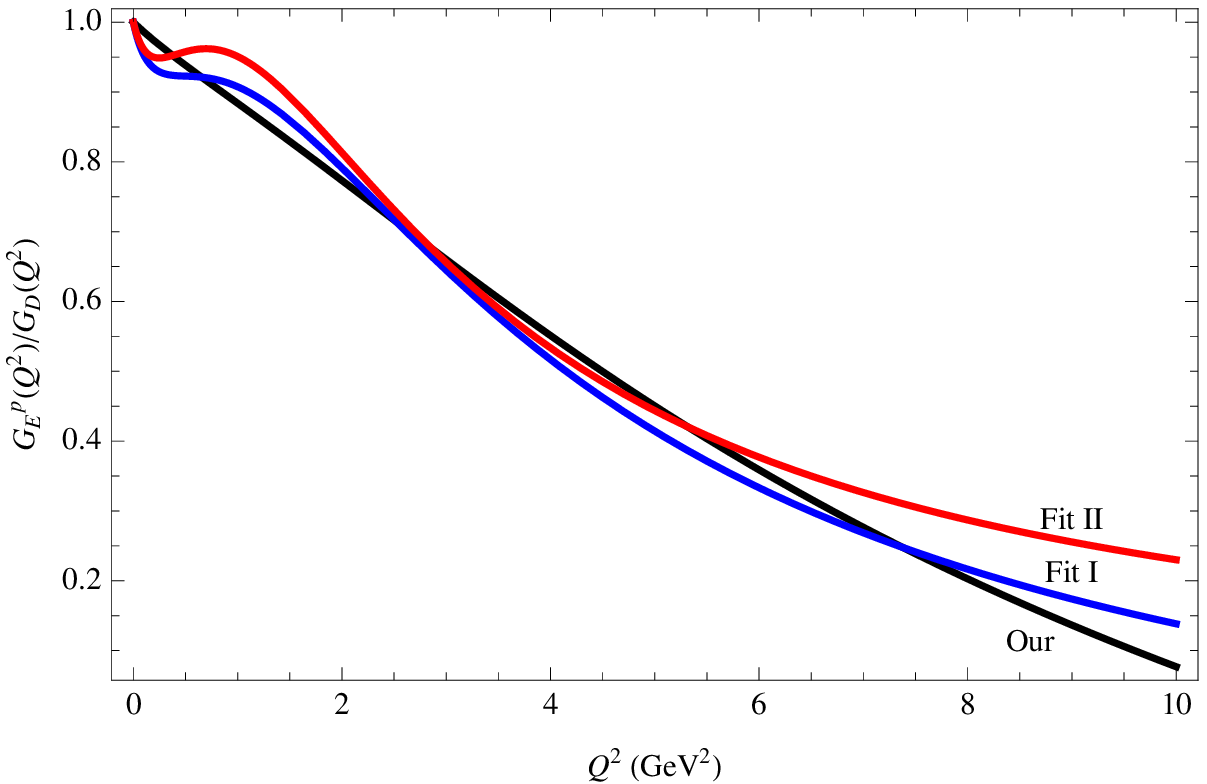,scale=.65}
\epsfig{figure=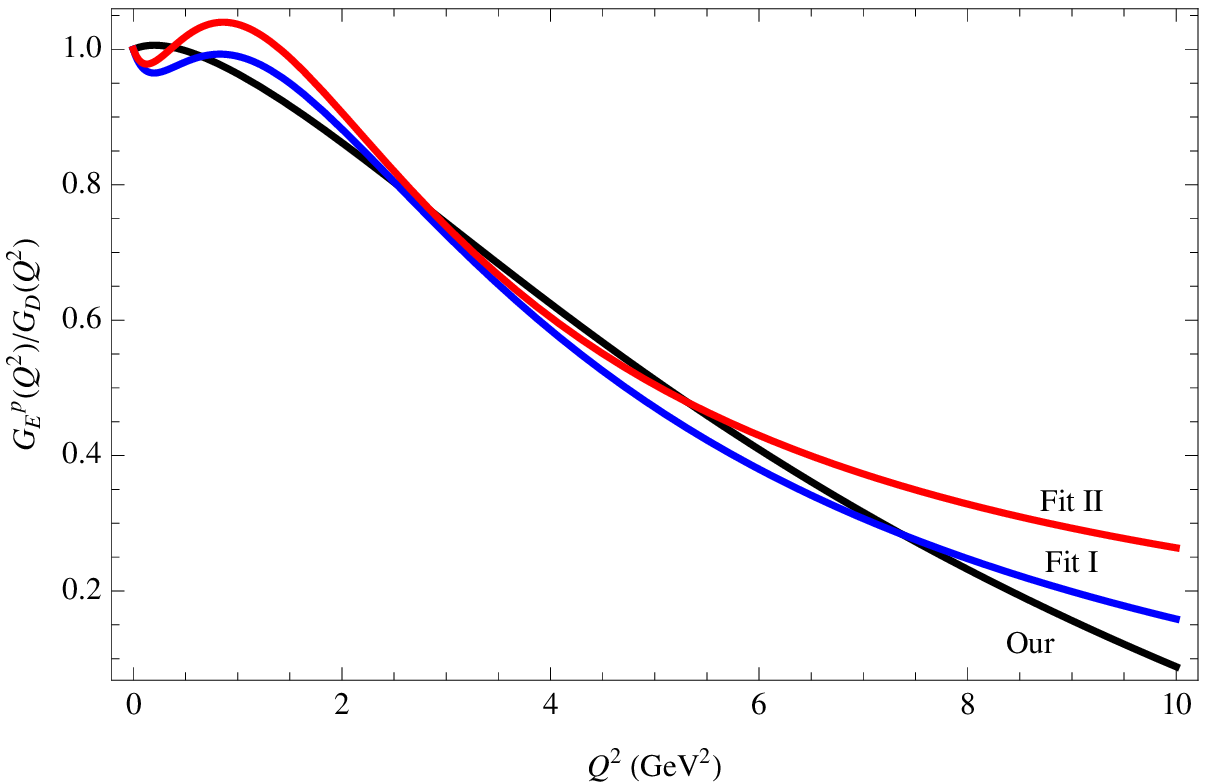,scale=.65}
\end{center}
\noindent
\caption{Ratio $G_E^p(Q^2)/G_D(Q^2)$ 
in comparison with global Fit I and Fit II 
for dipole scale parameter $\Lambda^2 = 0.71$ GeV$^2$ 
(left panel) and $\Lambda^2 = 0.66$ GeV$^2$ (right panel). 
\label{fig5}}
\begin{center}
\epsfig{figure=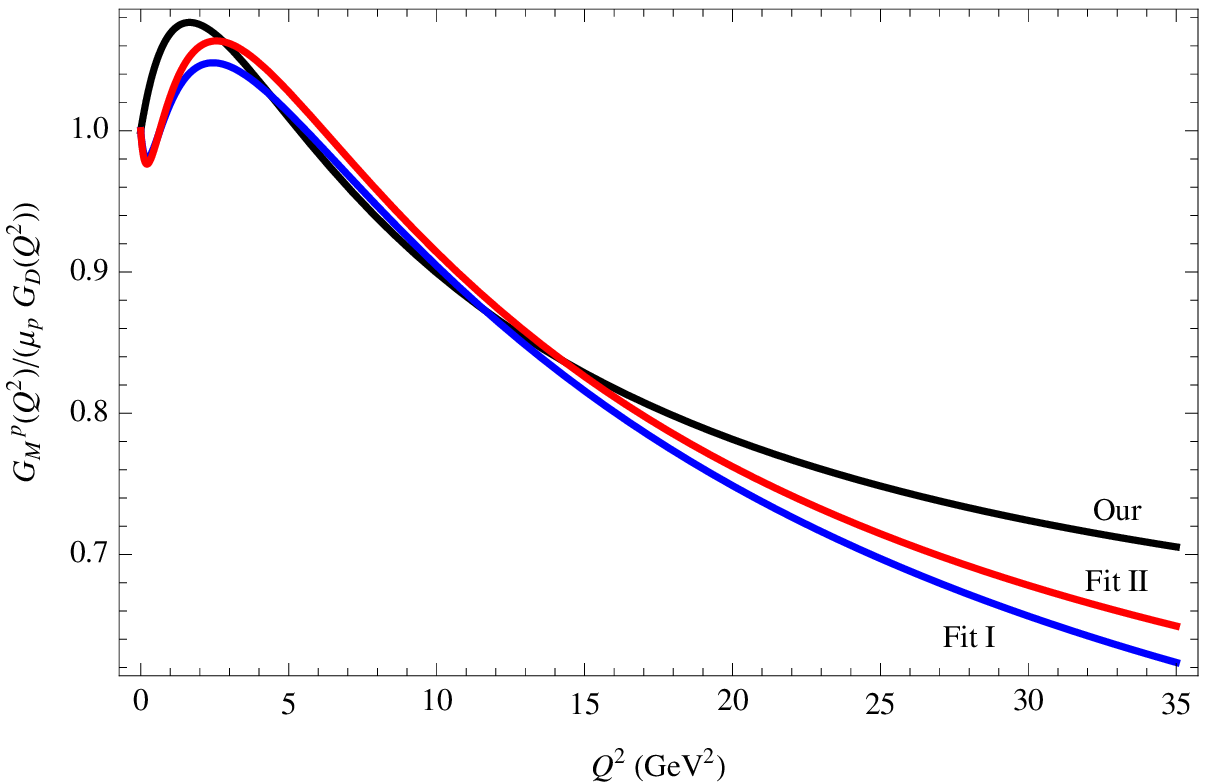,scale=.65}
\epsfig{figure=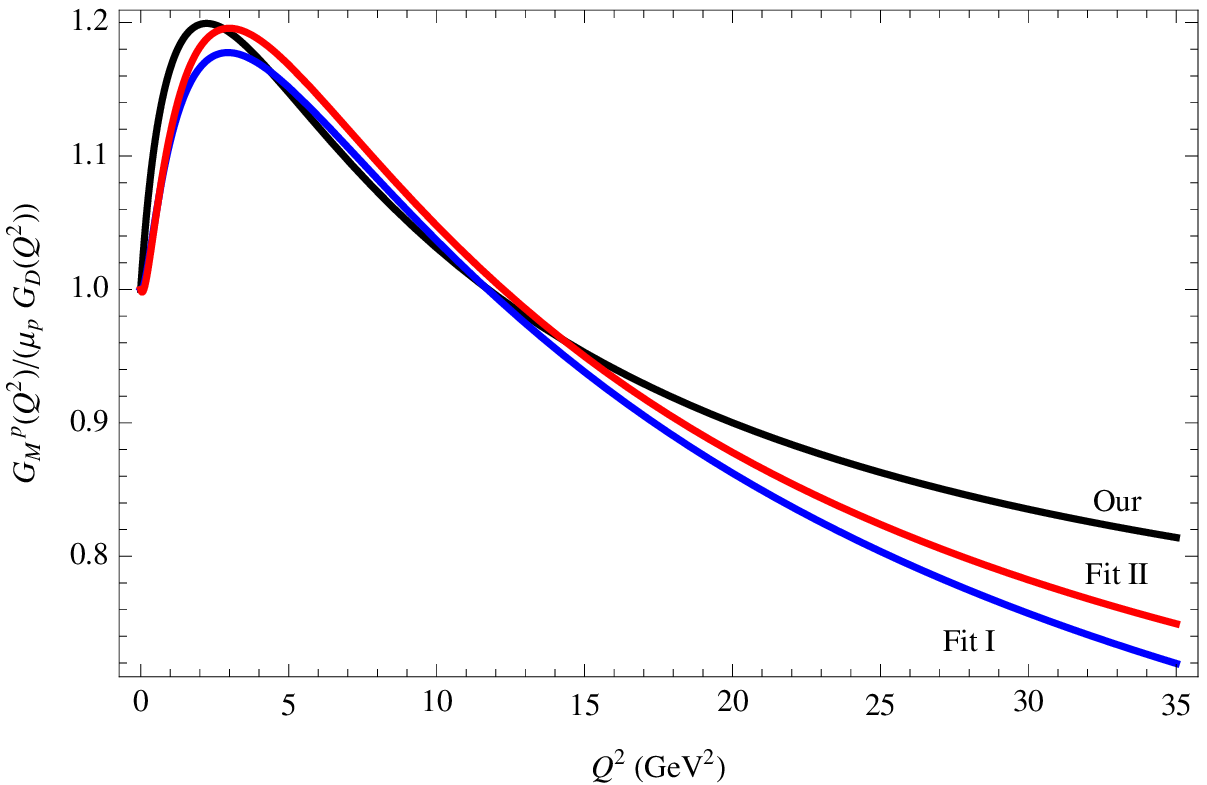,scale=.65}
\end{center}
\noindent
\caption{Ratio $G_M^p(Q^2)/(\mu_p G_D(Q^2))$ 
in comparison with global Fit I and Fit II 
for dipole scale parameter $\Lambda^2 = 0.71$ GeV$^2$ 
(left panel) and $\Lambda^2 = 0.66$ GeV$^2$ (right panel). 
\label{fig6}}
\end{figure}

\clearpage 

\begin{figure}
\begin{center}
\epsfig{figure=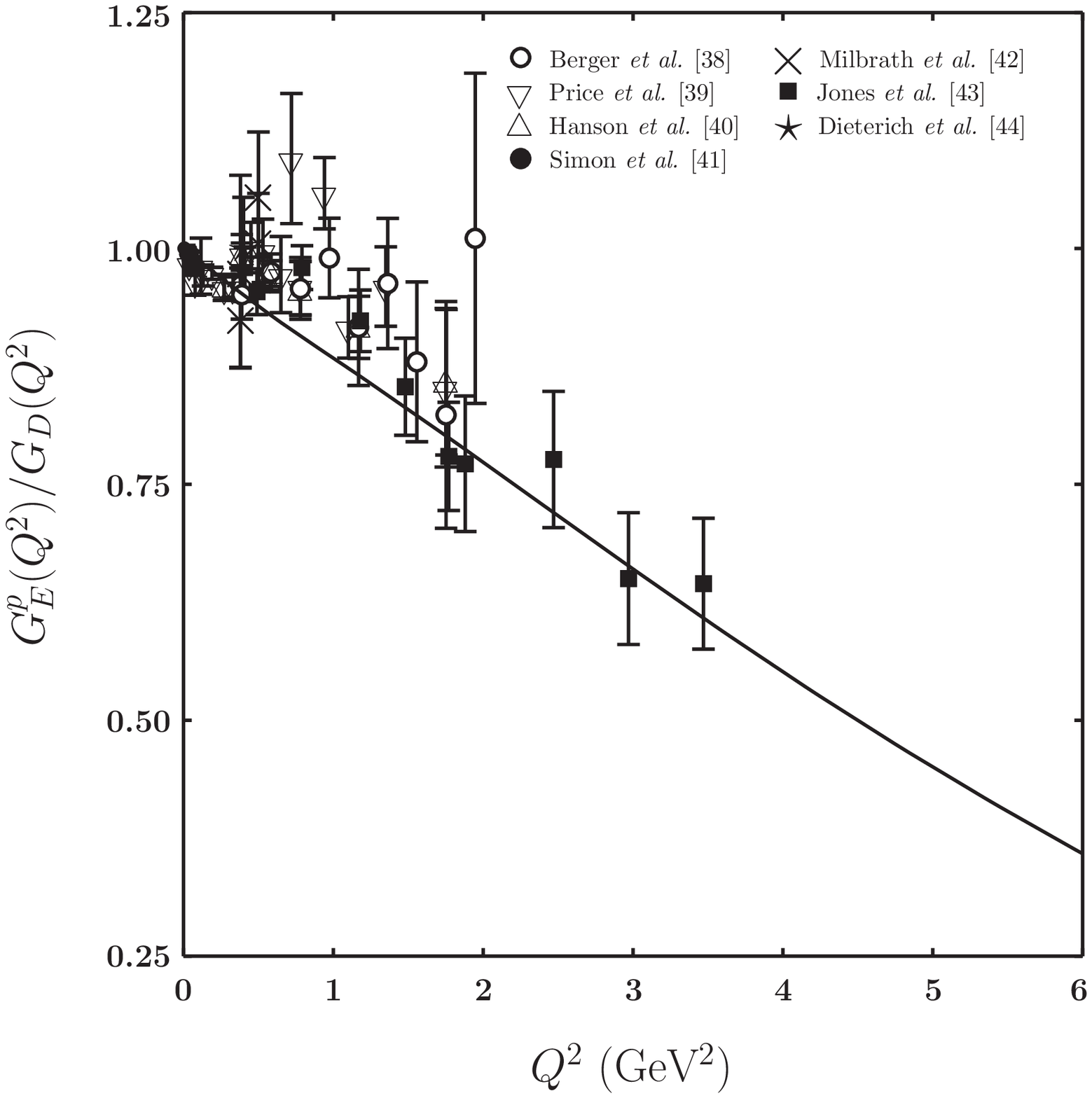,scale=.38}
\epsfig{figure=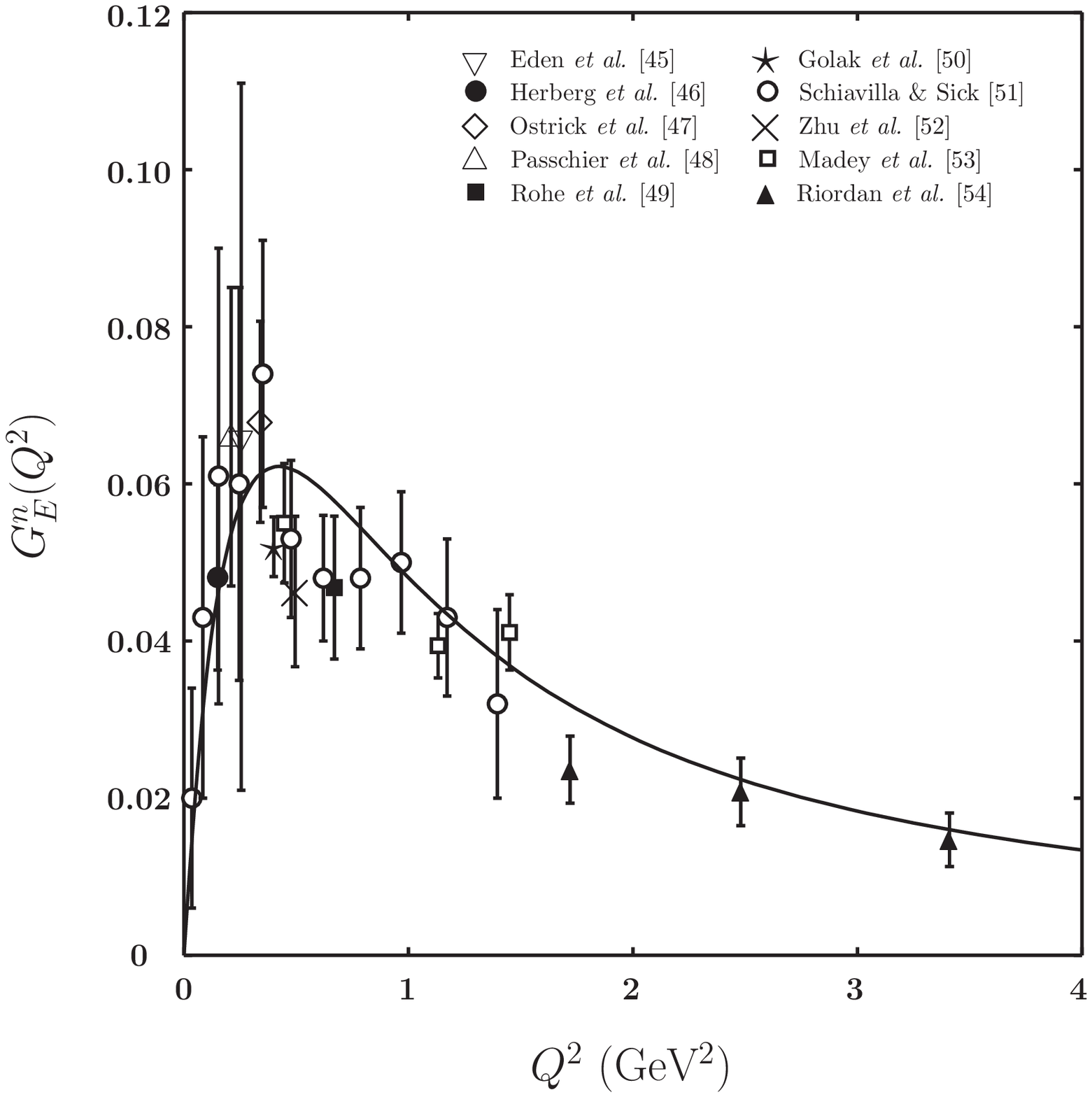,scale=.38}
\end{center}
\vspace*{-1.5cm}
\noindent
\caption{Ratio $G_E^p(Q^2)/G_D(Q^2)$ and 
charge neutron form factor $G_E^n(Q^2)$ 
in comparison with data.
\label{fig7}}
\begin{center}
\epsfig{figure=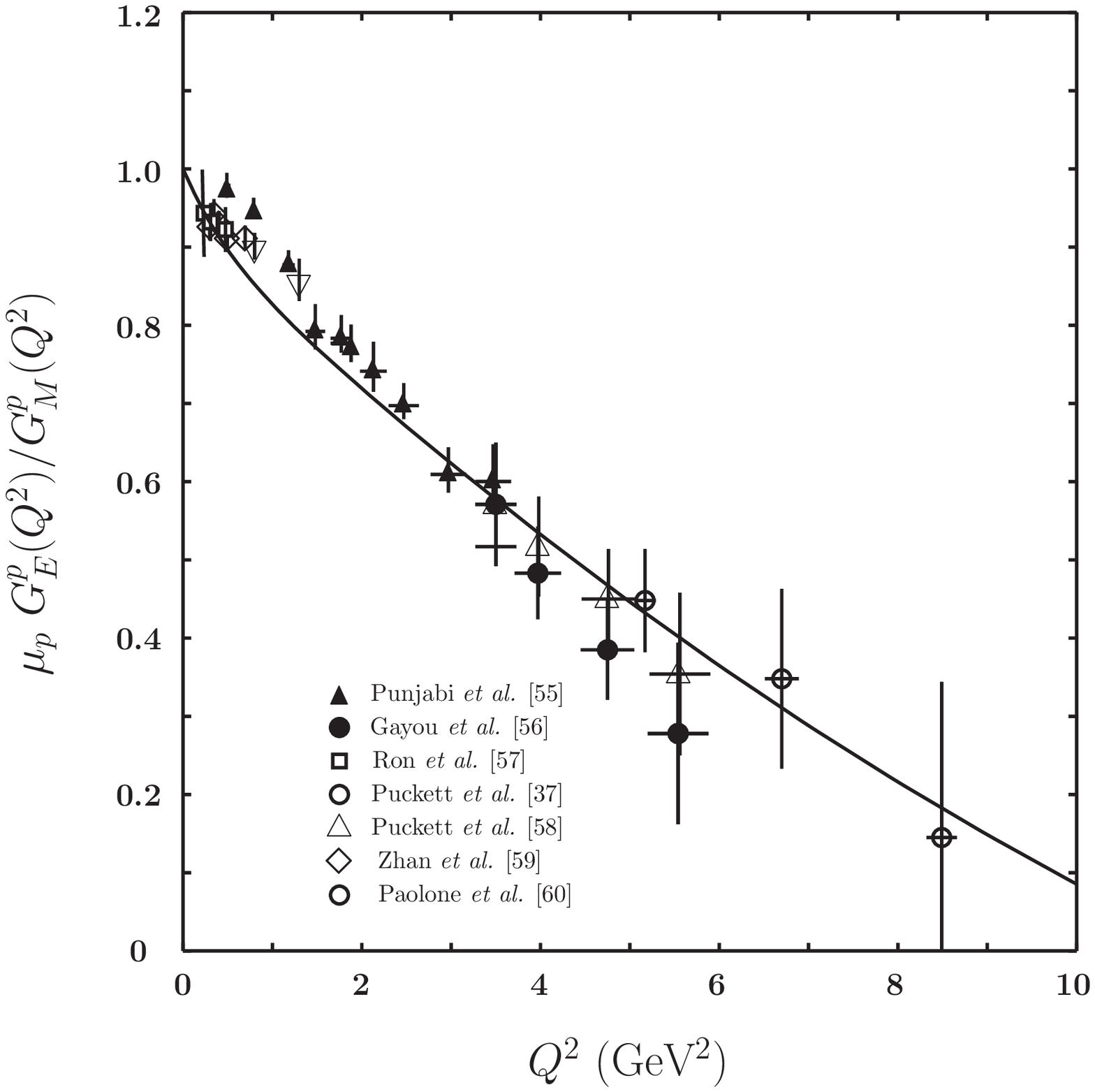,scale=.38}
\epsfig{figure=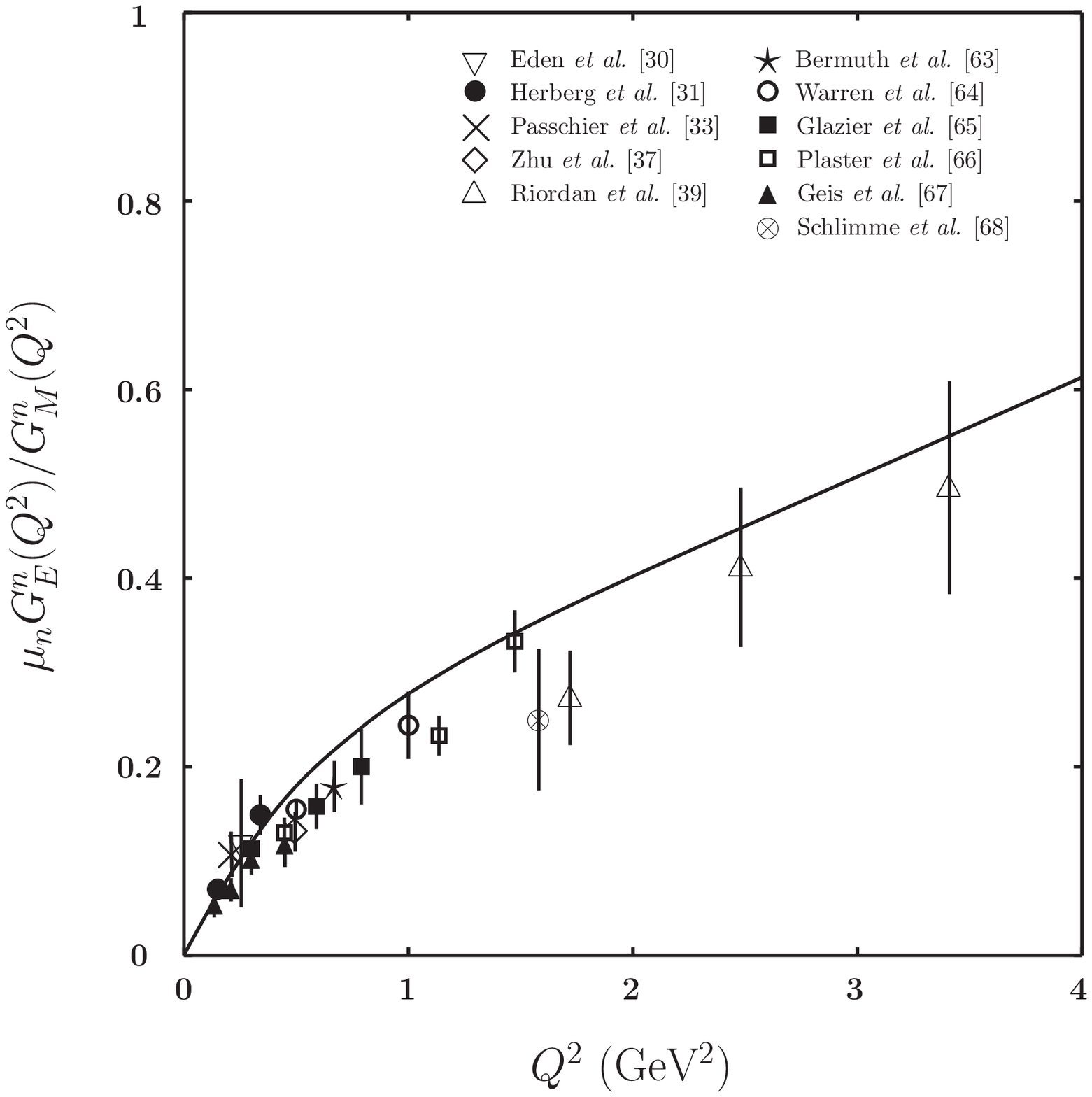,scale=.38}
\end{center}
\vspace*{-1.5cm}
\noindent
\caption{Ratios $\mu_p G_E^p(Q^2)/G_M^p(Q^2)$
and $\mu_n G_E^n(Q^2)/G_M^n(Q^2)$ 
in comparison with data.
\label{fig8}}
\begin{center}
\epsfig{figure=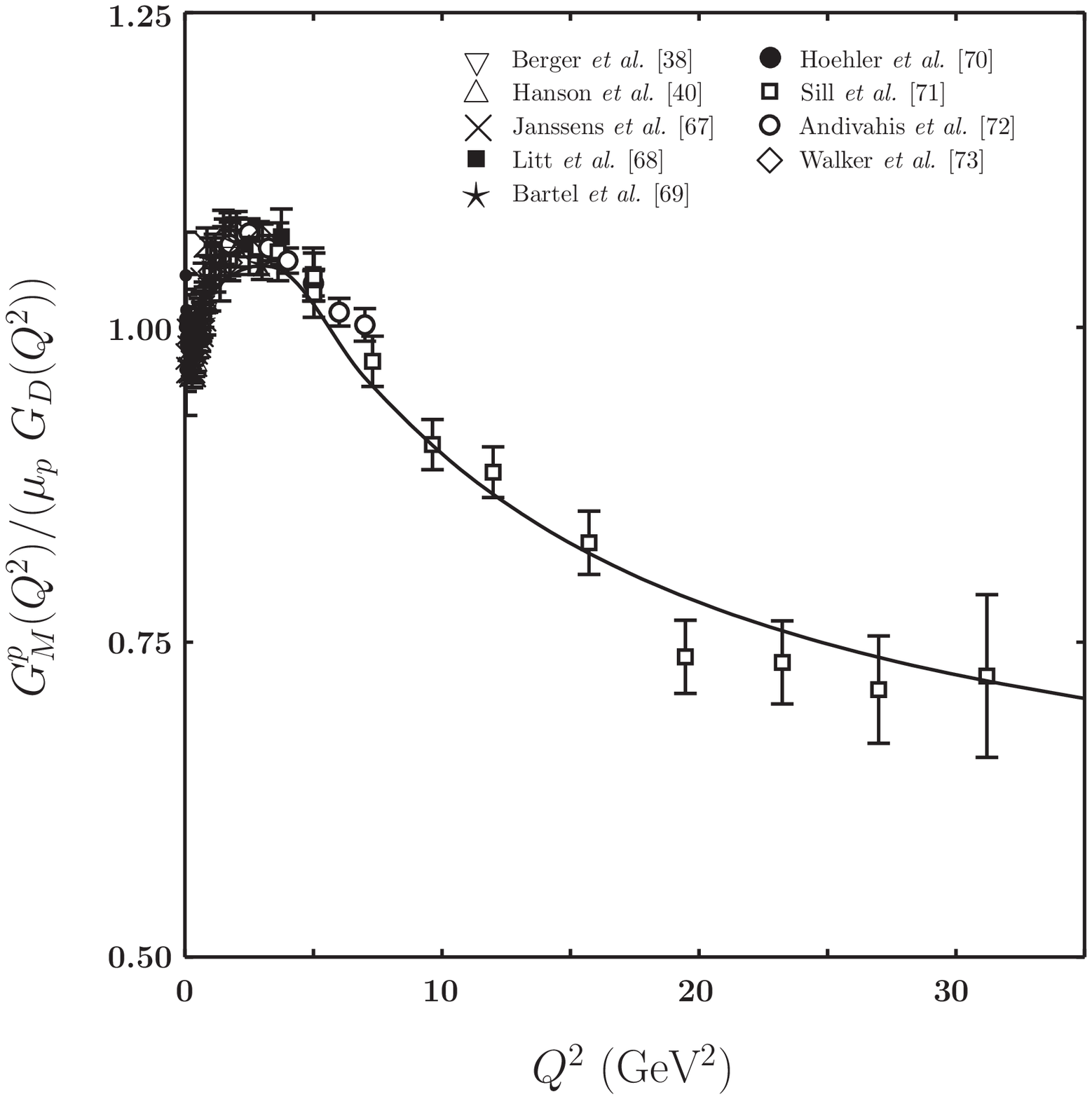,scale=.38}
\epsfig{figure=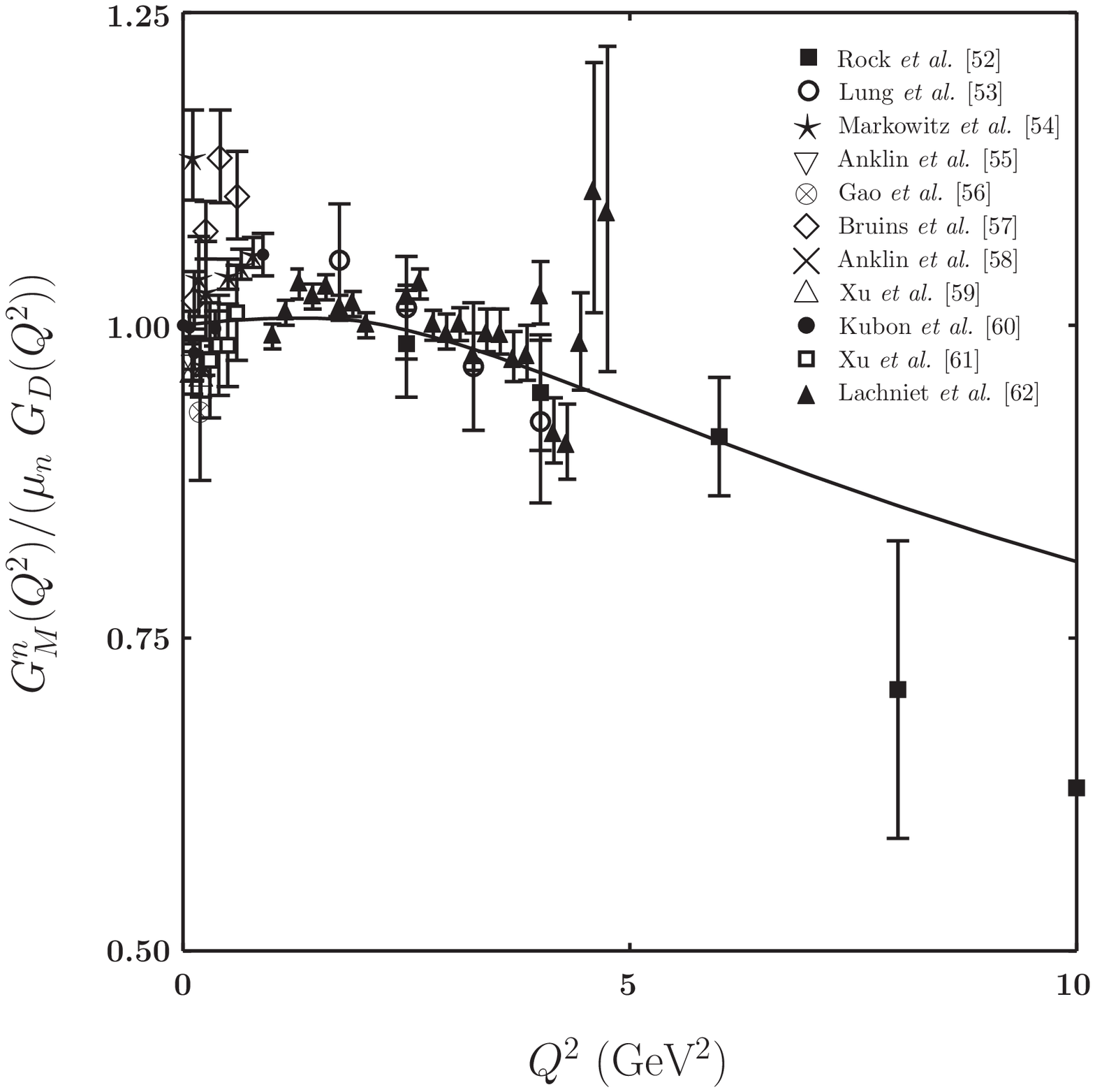,scale=.38}
\end{center}
\vspace*{-1.5cm}
\noindent
\caption{Ratios $G_M^p(Q^2)/(\mu_p G_D(Q^2))$.
and $G_M^n(Q^2)/(\mu_n G_D(Q^2))$ 
in comparison with data. 
\label{fig9}}
\end{figure}

\begin{figure}
\begin{center}
\epsfig{figure=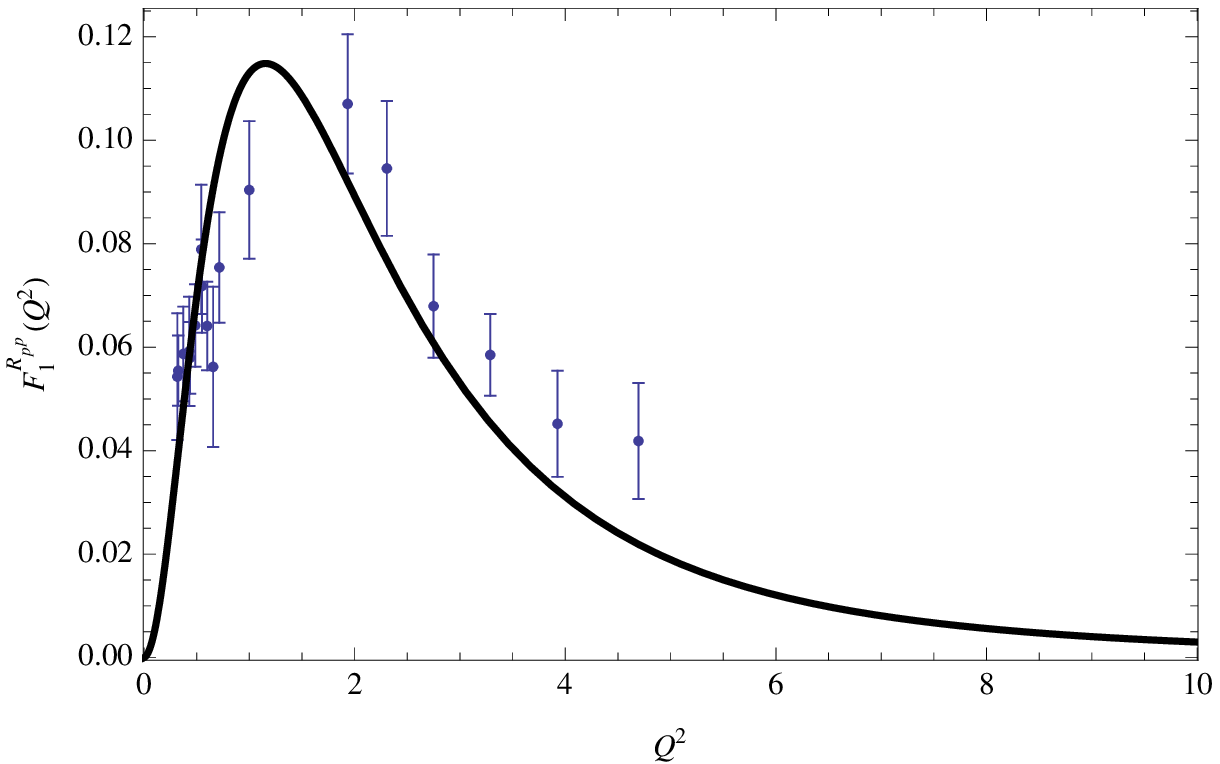,scale=.65}
\epsfig{figure=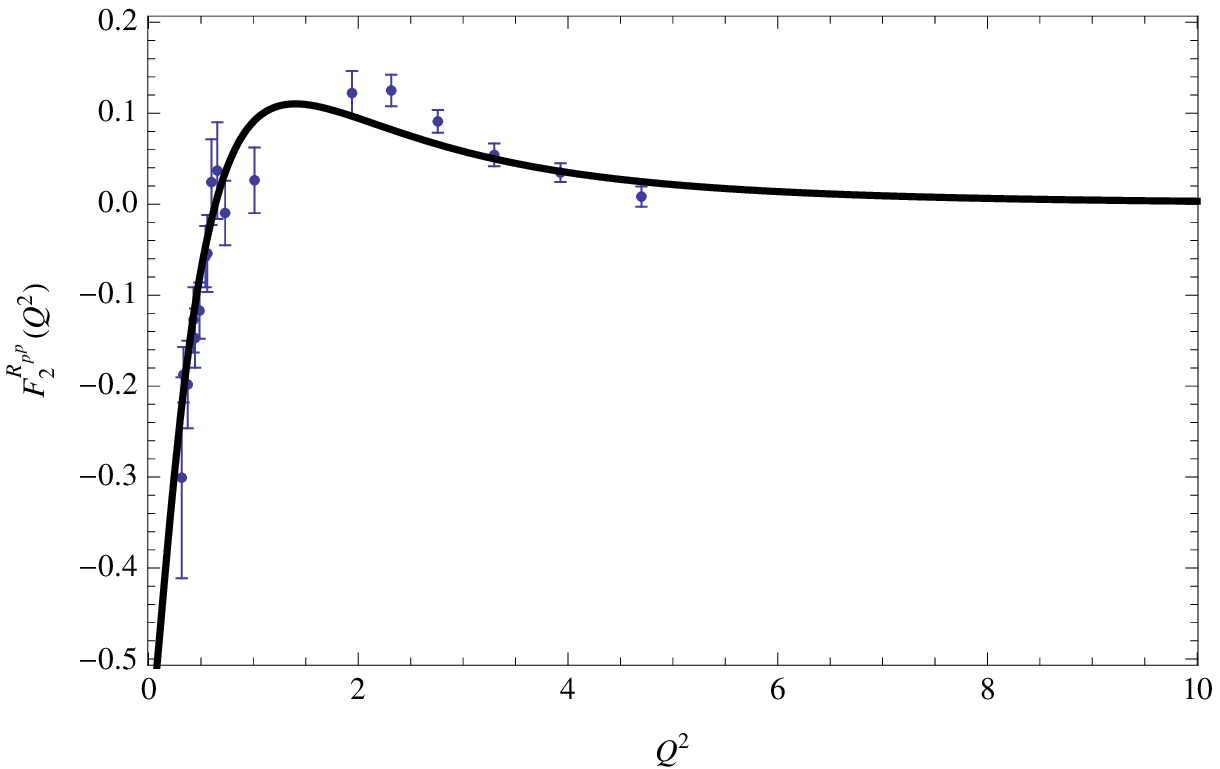,scale=.65}
\end{center}
\noindent
\caption{Roper-nucleon transition form factors 
$F_1^{{\cal R}_p p}(Q^2)$ and $F_2^{{\cal R}_p p}(Q^2)$ 
up to 10 GeV$^2$. 
\label{fig10}
}
\vspace*{.5cm}
\begin{center}
\epsfig{figure=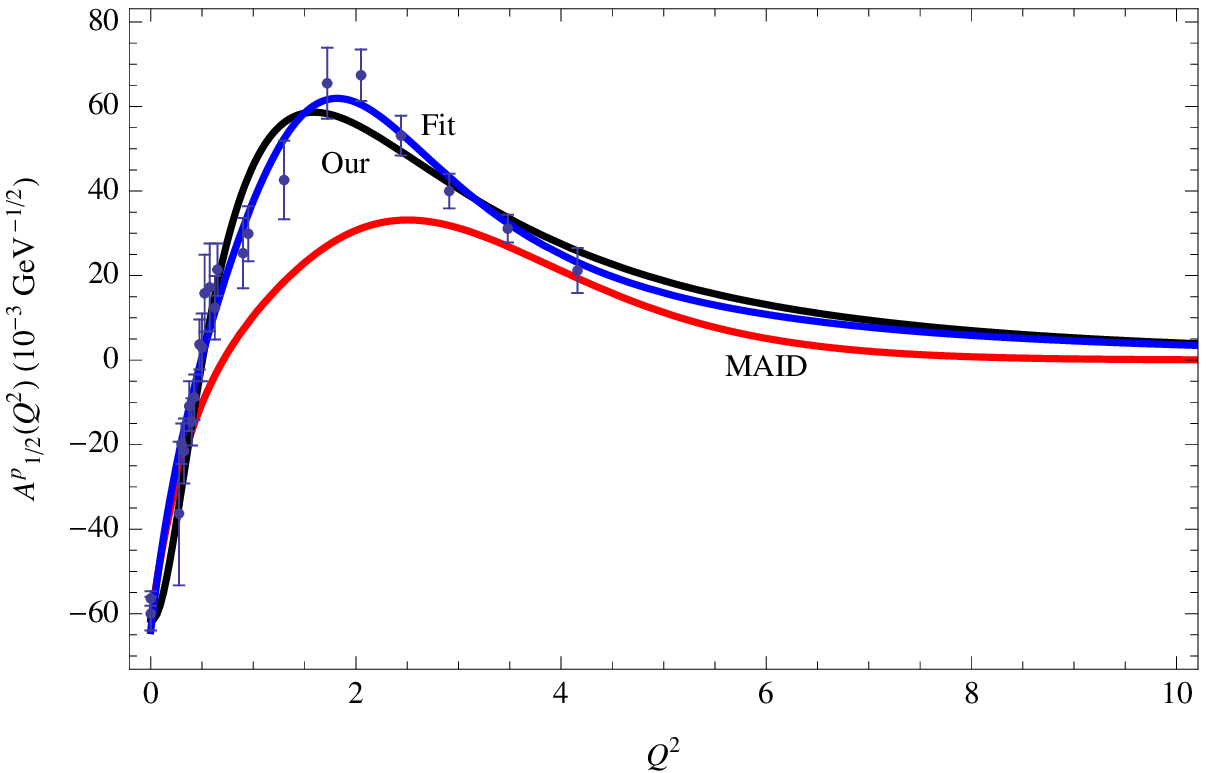,scale=.65}
\epsfig{figure=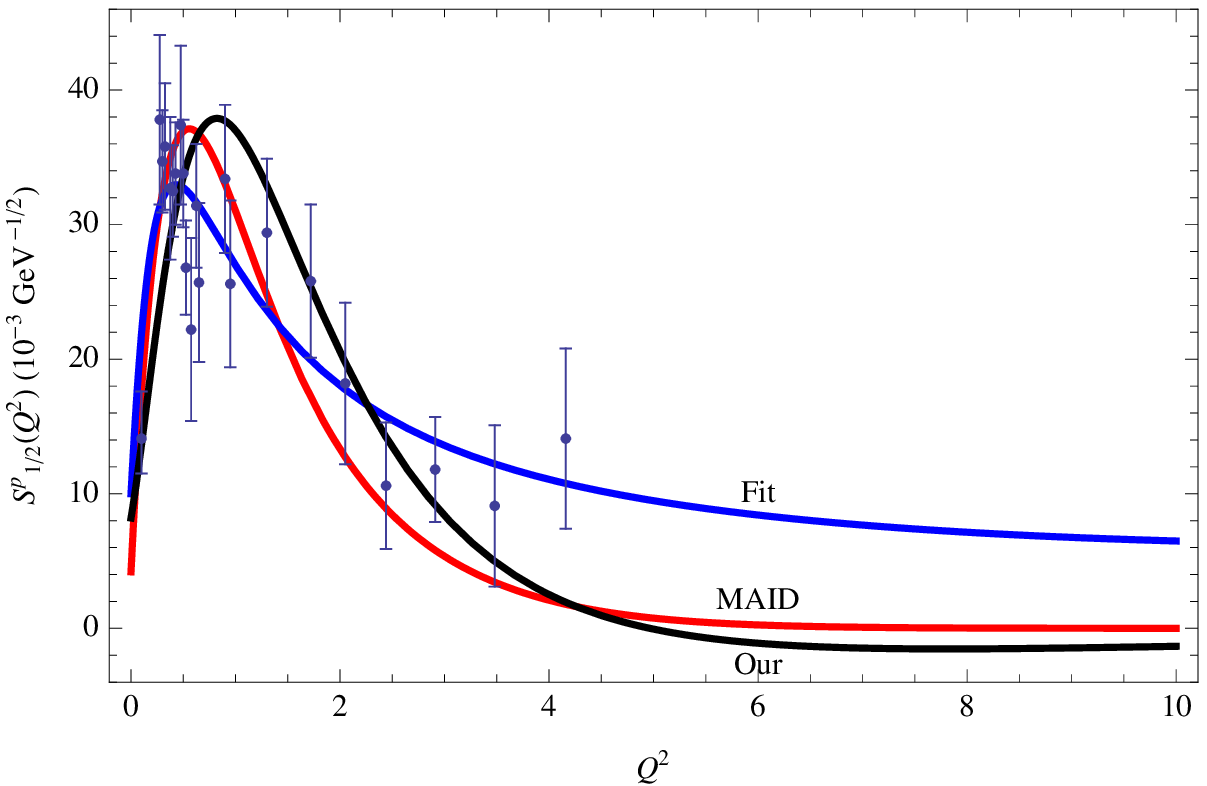,scale=.65}
\end{center}
\noindent
\caption{Helicity amplitudes $A_{1/2}^p(Q^2)$ 
and $S_{1/2}^p(Q^2)$ up to 10 GeV$^2$. 
\label{fig11}}

\end{figure}


\begin{thebibliography}{999}

\bibitem{Abidin:2009hr}
  Z.~Abidin and C.~E.~Carlson,
  Phys.\ Rev.\  D {\bf 79}, 115003 (2009).

\bibitem{Vega:2010ns}                                                          
  A.~Vega, I.~Schmidt, T.~Gutsche, and V.~E.~Lyubovitskij,
  Phys.\ Rev.\  D {\bf 83}, 036001 (2011).

\bibitem{Gutsche:2012bp} 
  T.~Gutsche, V.~E.~Lyubovitskij, I.~Schmidt, and A.~Vega,
  Phys.\ Rev.\ D {\bf 86}, 036007 (2012).

\bibitem{Brodsky:1973kr} 
  S.~J.~Brodsky and G.~R.~Farrar,
  Phys.\ Rev.\ Lett.\  {\bf 31}, 1153 (1973); 
  V.~A.~Matveev, R.~M.~Muradian and A.~N.~Tavkhelidze,
  Lett.\ Nuovo Cim.\  {\bf 7}, 719 (1973).  

\bibitem{Gutsche:2011vb}
  T.~Gutsche, V.~E.~Lyubovitskij, I.~Schmidt and A.~Vega,
  Phys.\ Rev.\ D {\bf 85}, 076003 (2012).

\bibitem{Brodsky:2014yha} 
  S.~J.~Brodsky, G.~F.~de Teramond, H.~G.~Dosch and J.~Erlich,
  Phys.\ Rept.\  {\bf 584}, 1 (2015).

\bibitem{Sufian:2016hwn} 
  R.~S.~Sufian, G.~F.~de Teramond, S.~J.~Brodsky, A.~Deur, and H.~G.~Dosch,
  Phys.\ Rev.\ D {\bf 95}, 014011 (2017). 

\bibitem{Chakrabarti:2013dda} 
  D.~Chakrabarti and C.~Mondal,
  Eur.\ Phys.\ J.\ C {\bf 73}, 2671 (2013).  

\bibitem{Gutsche:2013zia} 
  T.~Gutsche, V.~E.~Lyubovitskij, I.~Schmidt, and A.~Vega,
  Phys.\ Rev.\ D {\bf 89}, 054033 (2014), 
  Phys.\ Rev.\ D {\bf 92}, 019902(E) (2015).

\bibitem{Gutsche:2014yea} 
  T.~Gutsche, V.~E.~Lyubovitskij, I.~Schmidt, and A.~Vega,
  Phys.\ Rev.\ D {\bf 91}, 054028 (2015).

\bibitem{Gutsche:2016gcd} 
  T.~Gutsche, V.~E.~Lyubovitskij and I.~Schmidt,
  Eur.\ Phys.\ J.\ C {\bf 77}, 86 (2017).

\bibitem{DGLAP} 
  V.~N.~Gribov and L.~N.~Lipatov,
  Sov.\ J.\ Nucl.\ Phys.\  {\bf 15}, 438 (1972)
  [Yad.\ Fiz.\  {\bf 15}, 781 (1972)]; 
  V.~N.~Gribov and L.~N.~Lipatov,
  Sov.\ J.\ Nucl.\ Phys.\  {\bf 15}, 675 (1972)
  [Yad.\ Fiz.\  {\bf 15}, 1218 (1972)]; 
  G.~Altarelli and G.~Parisi,
  Nucl.\ Phys.\ B {\bf 126}, 298 (1977); 
  Y.~L.~Dokshitzer,
  Sov.\ Phys.\ JETP {\bf 46}, 641 (1977)
  [Zh.\ Eksp.\ Teor.\ Fiz.\  {\bf 73}, 1216 (1977)]. 

\bibitem{Brodsky:2016uln} 
  S.~J.~Brodsky, R.~F.~Lebed and V.~E.~Lyubovitskij,
  Phys.\ Lett.\ B {\bf 764}, 174 (2017).

\bibitem{deTeramond:2011qp}
  G.~F.~de Teramond and S.~J.~Brodsky,
  AIP Conf.\ Proc.\  {\bf 1432}, 168 (2012). 

\bibitem{Gutsche:2012wb} 
  T.~Gutsche, V.~E.~Lyubovitskij, I.~Schmidt, and A.~Vega,
  Phys.\ Rev.\ D {\bf 87}, 016017 (2013).

\bibitem{Ramalho:2017pyc} 
  G.~Ramalho and D.~Melnikov,
  arXiv:1703.03819 [hep-ph].

\bibitem{Ramalho:2017muv} 
  G.~Ramalho,
  Phys.\ Rev.\ D {\bf 96}, 054021 (2017). 
  
\bibitem{Aznauryan:2012ba}
  I.~G.~Aznauryan {\it et al.},
  Int.\ J.\ Mod.\ Phys.\ E {\bf 22}, 1330015 (2013).  

\bibitem{Obukhovsky:2011sc}
  I.~T.~Obukhovsky, A.~Faessler, D.~K.~Fedorov, T.~Gutsche,  
  and V.~E.~Lyubovitskij,
  Phys.\ Rev.\ D {\bf 84}, 014004 (2011). 
  
\bibitem{Aznauryan:2012ec}
  I.~G.~Aznauryan and V.~D.~Burkert,
  Phys.\ Rev.\ C {\bf 85}, 055202 (2012). 

\bibitem{Aznauryan:2009mx}
  I.~G.~Aznauryan {\it et al.}  (CLAS Collaboration),
  Phys.\ Rev.\ C {\bf 80}, 055203 (2009). 

\bibitem{Mokeev:2012sha}
  V.~I.~Mokeev {\it et al.}  (CLAS Collaboration),
  Phys.\ Rev.\ C {\bf 86}, 035203 (2012). 

\bibitem{Mokeev:2015lda} 
  V.~I.~Mokeev {\it et al.},
  Phys.\ Rev.\ C {\bf 93}, 025206 (2016).

\bibitem{Grigoryan:2007my}
  H.~R.~Grigoryan and A.~V.~Radyushkin,
  Phys.\ Rev.\  D {\bf 76}, 095007 (2007).

 \bibitem{Kadeer:2005aq}
  A.~Kadeer, J.~G.~K\"orner and U.~Moosbrugger,
  Eur.\ Phys.\ J.\  {\bf C59}, 27 (2009).

\bibitem{Faessler:2009xn}
  A.~Faessler, T.~Gutsche, M.~A.~Ivanov, J.~G.~K\"orner,  
  and V.~E.~Lyubovitskij,
  Phys.\ Rev.\ D {\bf 80}, 034025 (2009). 

\bibitem{Branz:2010pq}
  T.~Branz, A.~Faessler, T.~Gutsche, M.~A.~Ivanov,
  J.~G.~K\"orner, V.~E.~Lyubovitskij, and B.~Oexl,
  Phys.\ Rev.\ D {\bf 81}, 114036 (2010).  

\bibitem{Gutsche:2017wag} 
  T.~Gutsche, M.~A.~Ivanov, J.~G.~K\"orner, V.~E.~Lyubovitskij, 
  V.~V.~Lyubushkin, and P.~Santorelli,
  Phys.\ Rev.\ D {\bf 96}, 013003 (2017).

\bibitem{Weber:1989fv}
  H.~J.~Weber,
  Phys.\ Rev.\ C {\bf 41}, 2783 (1990).
  
\bibitem{Aznauryan:2007ja}
  I.~G.~Aznauryan,
  Phys.\ Rev.\ C {\bf 76}, 025212 (2007); 
  I.~G.~Aznauryan, V.~D.~Burkert, and T.~-S.~H.~Lee,
  arXiv:0810.0997 [nucl-th].

 \bibitem{Copley:1972tu}
  L.~A.~Copley, G.~Karl, and E.~Obryk,
  Phys.\ Rev.\ D {\bf 4}, 2844 (1971).

 \bibitem{Capstick:1994ne}
  S.~Capstick and B.~D.~Keister,
  Phys.\ Rev.\ D {\bf 51}, 3598 (1995). 

 \bibitem{Tiator:2008kd}
  L.~Tiator and M.~Vanderhaeghen,
  Phys.\ Lett.\ B {\bf 672}, 344 (2009). 

 \bibitem{PDG:2016}
 C.~ Patrignani {\it et al.}
 (Particle Data Group), Chin.\ Phys.\ C {\bf 40}, 100001 (2016).

 \bibitem{Cates:2011pz}
  G.~D.~Cates, C.~W.~de Jager, S.~Riordan, and B.~Wojtsekhowski,
  Phys.\ Rev.\ Lett.\  {\bf 106}, 252003 (2011).

 \bibitem{Diehl:2013xca}
  M.~Diehl and P.~Kroll,
  Eur.\ Phys.\ J.\ C {\bf 73}, 2397 (2013);
  M.~Diehl,
  Nucl.\ Phys.\ Proc.\ Suppl.\  {\bf 161}, 49 (2006).

 \bibitem{Puckett:2017flj} 
  A.~J.~R.~Puckett {\it et al.},
  Phys.\ Rev.\ C {\bf 96}, 055203 (2017)

\bibitem{berger71}
  C.~Berger, V.~Burkert, G.~Knop, B.~Langenbeck and K.~Rith,
  Phys.\ Lett.\ B {\bf 35}, 87 (1971).

\bibitem{price71}
  L.~E.~Price {\it et al.},
  Phys.\ Rev.\ D {\bf 4}, 45 (1971).

\bibitem{hanson73}
  K.~M.~Hanson {\it et al.},
  Phys.\ Rev.\ D {\bf 8}, 753 (1973).

\bibitem{simon80}
  G.~G.~Simon, C.~Schmitt, F.~Borkowski and V.~H.~Walther,
  Nucl.\ Phys.\ A {\bf 333}, 381 (1980).

\bibitem{milbrath98}
 B.~D.~Milbrath {\it et al.} (Bates FPP Collaboration),
 Phys.\ Rev.\ Lett.\  {\bf 80}, 452 (1998);
                      {\bf 82}, 2221(E) (1999).

\bibitem{jones00}
  M.~K.~Jones {\it et al.} (Jefferson Lab Hall A Collaboration),
  Phys.\ Rev.\ Lett.\  {\bf 84}, 1398 (2000).

\bibitem{dieterich01}
 S. Dieterich {\it et al.},
  Phys.\ Lett.\ B {\bf 500}, 47 (2001).

\bibitem{eden94}
 T.~Eden {\it et al.},
Phys.\ Rev.\  C {\bf 50}, R1749 (1994).

\bibitem{herberg99}
 C.~Herberg {\it et al.},
  Eur.\ Phys.\ J.\ A {\bf 5}, 131 (1999).

\bibitem{ostrick99}
 M.~Ostrick {\it et al.},
  Phys.\ Rev.\ Lett.\  {\bf 83}, 276 (1999).

\bibitem{passchier99}
 I.~Passchier {\it et al.},
  Phys.\ Rev.\ Lett.\  {\bf 82}, 4988 (1999).

\bibitem{rohe99}
 D. Rohe {\it et al.},
  Phys.\ Rev.\ Lett.\  {\bf 83}, 4257 (1999).

\bibitem{golak01}
J.~Golak, G.~Ziemer, H.~Kamada, H.~Witala, and W.~Gl\"ockle,
  Phys.\ Rev.\ C {\bf 63}, 034006 (2001).

\bibitem{schiavilla01}
R.~Schiavilla and I.~Sick,
  Phys.\ Rev.\ C {\bf 64}, 041002 (2001).

\bibitem{zhu01}
 H.~Zhu {\it et al.} (Jefferson Lab E93-026 Collaboration),
  Phys.\ Rev.\ Lett.\  {\bf 87}, 081801 (2001).

\bibitem{madey03}
 R. Madey {\it et al.} (Jefferson Lab E93-038 Collaboration),
  Phys.\ Rev.\ Lett.\  {\bf 91}, 122002 (2003).

\bibitem{Riordan:2010id}
  S.~Riordan {\it et al.},
  Phys.\ Rev.\ Lett.\  {\bf 105}, 262302 (2010).

\bibitem{punjabi05}
   V.~Punjabi {\it et al.},
  Phys.\ Rev.\ C {\bf 71}, 055202 (2005);
  C {\bf 71}, 069902(E) (2005).

\bibitem{gayou02}
 O.~Gayou {\it et al.}  (Jefferson Lab Hall A Collaboration),
  Phys.\ Rev.\ Lett.\  {\bf 88}, 092301 (2002).

\bibitem{ron11}
 G.~Ron {\it et al.} (JLab Hall A Collaboration),
  Phys.\ Rev.\ C {\bf 84}, 055204 (2011).

\bibitem{puckett12}
 A.~J.~R.~Puckett {\it et al.},
  Phys.\ Rev.\ C {\bf 85}, 045203 (2012).

\bibitem{Zhan:2011ji} 
  X.~Zhan {\it et al.},
  Phys.\ Lett.\ B {\bf 705}, 59 (2011). 

\bibitem{Paolone:2010qc} 
  M.~Paolone {\it et al.},
  Phys.\ Rev.\ Lett.\  {\bf 105}, 072001 (2010). 

\bibitem{Bermuth:2003qh}
  J.~Bermuth {\it et al.},
  Phys.\ Lett.\ B {\bf 564}, 199 (2003).

\bibitem{Warren:2003ma}
  G.~Warren {\it et al.}  (Jefferson Laboratory E93-026 Collaboration),
  Phys.\ Rev.\ Lett.\  {\bf 92}, 042301 (2004).

\bibitem{Glazier:2004ny}
  D.~I.~Glazier {\it et al.},
  Eur.\ Phys.\ J.\ A {\bf 24}, 101 (2005)

\bibitem{Plaster:2005cx}
  B.~Plaster {\it et al.}  (Jefferson Laboratory E93-038 Collaboration),
  Phys.\ Rev.\  C {\bf 73}, 025205 (2006).

\bibitem{Geis:2008aa}
  E.~Geis {\it et al.}  (BLAST Collaboration),
  Phys.\ Rev.\ Lett.\  {\bf 101}, 042501 (2008).

\bibitem{Schlimme:2013eoz}
  B.~S.~Schlimme {\it et al.},
  Phys.\ Rev.\ Lett.\  {\bf 111}, 132504 (2013).

\bibitem{Janssens:1966}
  T.~Janssens, R.~Hofstadter, E.~B.~Hughes, and M.~R.~Yearian,
  Phys.\ Rev.\ {\bf 142}, 922 (1966).

\bibitem{Litt:1969my}
  J.~Litt {\it et al.},
  Phys.\ Lett.\ B {\bf 31}, 40 (1970).

\bibitem{bartel73}
  W. Bartel {\it et al.},
  Nucl.\ Phys.\  {\bf B58}, 429 (1973).

\bibitem{Hohler:1976ax}
  G.~Hohler {\it et al.},
  Nucl.\ Phys.\ B {\bf 114}, 505 (1976).

\bibitem{Sill:1992qw}
  A.~F.~Sill {\it et al.},
  Phys.\ Rev.\ D {\bf 48}, 29 (1993).

\bibitem{Andivahis:1994rq}
  L.~Andivahis {\it et al.},
  Phys.\ Rev.\ D {\bf 50}, 5491 (1994).

\bibitem{Walker:1993vj}
  R.~C.~Walker {\it et al.},
  Phys.\ Rev.\ D {\bf 49} 5671 (1994).

\bibitem{Rock:1982gf}
  S.~Rock {\it et al.},
  Phys.\ Rev.\ Lett.\  {\bf 49}, 1139 (1982).

\bibitem{Lung:1992bu}
  A.~Lung {\it et al.},
  Phys.\ Rev.\ Lett.\  {\bf 70}, 718 (1993).

\bibitem{Markowitz:1993hx}
  P.~Markowitz {\it et al.},
  Phys.\ Rev.\ C {\bf 48}, R5 (1993).

\bibitem{Anklin:1994ae}
  H.~Anklin {\it et al.},
  Phys.\ Lett.\ B {\bf 336}, 313 (1994).


\bibitem{Gao:1994ud}
  H.~Gao {\it et al.},
  Phys.\ Rev.\ C {\bf 50}, R546 (1994).

\bibitem{Bruins:1995ns}
  E.~E.~W.~Bruins {\it et al.},
  Phys.\ Rev.\ Lett.\  {\bf 75}, 21 (1995).

\bibitem{Anklin:1998ae}
  H.~Anklin {\it et al.},
  Phys.\ Lett.\ B {\bf 428}, 248 (1998).

\bibitem{xu00}
  W.~Xu {\it et al.}
  Phys.\ Rev.\ Lett.\  {\bf 85}, 2900 (2000).

\bibitem{Kubon:2001rj}
  G.~Kubon {\it et al.},
  Phys.\ Lett.\ B {\bf 524}, 26 (2002).

\bibitem{Xu:2002xc}
  W.~Xu {\it et al.} (Jefferson Lab E95-001 Collaboration),
  Phys.\ Rev.\ C {\bf 67}, 012201 (2003).

\bibitem{Lachniet:2008qf}
  V.~J.~Lachniet {\it et al.} (CLAS Collaboration),
  Phys.\ Rev.\ Lett.\  {\bf 102}, 192001 (2009).

\bibitem{Stajner:2017fmh} 
  S.~Stajner {\it et al.},
  Phys.\ Rev.\ Lett.\  {\bf 119}, 022001 (2017).

\bibitem{Drechsel:2007if} 
  D.~Drechsel, S.~S.~Kamalov and L.~Tiator,
  Eur.\ Phys.\ J.\ A {\bf 34}, 69 (2007). 
 
\end{thebibliography}
\end{document}